# Network MIMO with Linear Zero-Forcing Beamforming:
# Large System Analysis, Impact of Channel Estimation and Reduced-Complexity Scheduling

Hoon Huh, *Member, IEEE*, Antonia M. Tulino, *Senior Member, IEEE*, and Giuseppe Caire, *Fellow, IEEE*

**Abstract**

We consider the downlink of a multi-cell system with multi-antenna base stations and single-antenna user terminals, arbitrary base station cooperation clusters, distance-dependent propagation pathloss, and general "fairness" requirements. Base stations in the same cooperation cluster employ *joint transmission* with linear zero-forcing beamforming, subject to sum or per-base station power constraints. Inter-cluster interference is treated as noise at the user terminals. Analytic expressions for the system spectral efficiency are found in the large-system limit where both the numbers of users and antennas per base station tend to infinity with a given ratio. In particular, for the per-base station power constraint, we find new results in random matrix theory, yielding the squared Frobenius norm of submatrices of the Moore-Penrose pseudo-inverse for the structured non-i.i.d. channel matrix resulting from the cooperation cluster, user distribution, and path-loss coefficients. The analysis is extended to the case of non-ideal *Channel State Information at the Transmitters* (CSIT) obtained through explicit downlink channel training and uplink feedback. Specifically, our results illuminate the trade-off between the benefit of a larger number of cooperating antennas and the cost of estimating higher-dimensional channel vectors. Furthermore, our analysis leads to a new simplified downlink scheduling scheme that pre-selects the users according to probabilities obtained from the large-system results, depending on the desired fairness criterion. The proposed scheme performs close to the optimal (finite-dimensional) opportunistic user selection while requiring significantly less channel state feedback, since only a small fraction of pre-selected users must feed back their channel state information.

## I. INTRODUCTION

The next generation of wireless communication systems (e.g., 802.16m [1], LTE-Advanced [2]) consider multiuser MIMO (MU-MIMO) as one of the core technologies. A considerable research effort has been dedicated to the performance evaluation of MU-MIMO systems in realistic cellular environments [3]–[5]. In a single-cell setting with perfect Channel State Information at the Transmitter (CSIT), the system reduces to a vector Gaussian broadcast channel whose capacity region is completely characterized [6]–[10]. However, in a multi-cell scenario, we are in the presence of a vector Gaussian broadcast and interference channel, which is not yet fully understood in an information theoretic sense.

A simple and practical approach consists of treating Inter-Cell Interference (ICI) as noise. In this case, ICI may significantly limit the system capacity. A variety of *inter-cell cooperation* schemes have been proposed to mitigate ICI, ranging from a fully cooperative network MIMO [11]–[14] to partially coordinated beamforming [15]–[17]. In this work, we focus on the network MIMO approach with limited cooperation, where clusters of cooperating base stations (BSs) act as a single distributed MIMO transmitter and interference from other clusters is treated as noise.

In a cellular environment, the received signal power is a polynomially decreasing function of the distance between transmitter and receiver, with a dynamic range typically larger than 30 dB [18]. Thus, users close to the cell (or cluster of coordinated cells) boundary experience strong inter-cell interference, whereas the desired signal is relatively weak. These "edge" users cannot be just *ignored* by the system. For example, maximizing the system sum-rate leads in general to a very unfair operating point, where the system resources are concentrated on users near the cell (or cluster) center. In contrast, *fairness scheduling* has been proposed and widely studied in order to achieve a desirable balance between sum-rate and fairness (see for example [19]–[21] and references therein). Fairness scheduling can be systematically implemented in the framework of *stochastic network optimization* [20]. Such fairness scheduling algorithms dynamically allocate the system resources on a slot-by-slot basis, such that the long-term average (or "ergodic") user rate point maximizes some suitable concave and componentwise increasing *network utility function* [22]. While the analytical characterization of the optimal ergodic rate point for a given network utility function may be hopelessly complicated in a realistic scenario, the system performance has been evaluated so far through computationally very intensive Monte Carlo simulation [3]–[5], [13], [14], [23]–[26], where the actual scheduling algorithm evolves in time and the ergodic rates are computed as time averages.

The capacity of a multi-cell network MIMO system under fairness criteria was evaluated in the large-system limit in [27], assuming ideal channel state knowledge and the Gaussian *Dirty Paper Coding* (DPC) transmission strategy. In [27], the asymptotic analysis method based on large random matrix results demonstrated its effectiveness by the comparison with finite-dimensional Monte Carlo simulation. In this work, we apply a similar approach to *Linear Zero-forcing Beamforming* (LZFB). It turns out that the analysis in this case is significantly more complicated, in particular, in order to take into account per-BS power constraint. In this paper we also extend our analysis to the case where the CSIT is obtained through explicit training and MMSE estimation. In these conditions, we obtain a lower bound on the achievable ergodic rates (referred to as "throughput" in the following), that takes into account the overhead due to training-based channel estimation. Several novel and important aspects are illuminated in this paper. Specifically: 1) As in [27], our analysis allows precise performance evaluation of systems for which brute-force Monte Carlo simulation would be very demanding. 2) By including the effect of training and channel estimation, we can investigate the tradeoff between ICI reduction owing to BS cooperation and the cost of estimating larger and larger dimensional channels. Unlike previous results that assumed ideal CSIT at no cost [24], [25], [27], we observe that there exist an optimal cooperation cluster size that depends on the channel coherence time and bandwidth, beyond which cooperative network MIMO is not convenient, consistently with the finite-dimensional simulation findings of [26], [28]. 3) We provide novel results in random matrix theory, in particular, related to the evaluation of the coefficients appearing in the per-BS power constraint. 4) We use our asymptotic analysis in order to design a *probabilistic scheduling* algorithm that randomly pre-selects the users with assigned probabilities obtained from the large-system results, and therefore requires much less CSIT feedback than the standard opportunistic scheduling scheme based on channel-driven user selection.

This last point deserves some remarks, since for the first time (to the authors' knowledge) asymptotic results are used not only for performance *analysis* but also for *system design* in network MIMO. [1] The standard approach to scheduling for downlink beamforming consists of having a large number of users feeding back their CSIT and selecting a subset of users with cardinality not larger than the number of jointly coordinated transmit antennas, such that the channel vectors of the selected users have both large norm and are mutually approximately orthogonal [31], [32]. This *multiuser diversity* selection, combined with LZFB precoding, is shown to attain the same performance as Gaussian DPC in the limit of a large number of users and fixed number of transmit antennas. However, in this limit, the throughput per user vanishes as $O(\frac{\log \log n}{n})$ where $n$ is the number of users. Therefore, a more meaningful regime is one in which the number of users is proportional to the number of antennas, yielding constant throughput per user. This is in fact the large-system regime investigated in this paper. Comparing the results of our asymptotic analysis with the Monte Carlo simulation of finite dimensional systems, including user selection as said before, we notice that multiuser diversity yields larger throughput per user for low-dimensional systems, but this gain reduces as the system dimension grows. This is a manifestation of the "channel hardening effect" noticed in [33], and agrees with the theoretical findings in [34], showing that the probability of finding a subset of approximately orthogonal users vanishes as the system dimension increases. Hence, as the system becomes large, there is diminishing return in selecting users from a large set. In contrast, the cost of CSIT feedback grows at least *linearly* with the number of users feeding back their estimated CSIT. Therefore, we advocate a probabilistic scheduling algorithm for which users are pre-selected at random using the probabilities derived from our large-system analysis, reflecting the desired fairness criterion, and only the selected users are required to feed back their CSIT. The performance of this scheme are shown to be close to the much more costly full user selection scheme, and become closer and closer as the system dimension increases (again, by the large-system limit and channel hardening effect).

---

[1] In the unrelated context of multiuser detection, asymptotic large-system results were used to design low-complexity linear multiuser detectors based on the polynomial approximations, where the polynomial coefficients were obtained from large random matrix theory [29], [30].

In comparison with concurrent existing literature, we notice that the LZFB MU-MIMO performance analysis with non-ideal CSIT was extensively studied in the finite-dimensional case (see for example [35]–[37]) and in the large-system limit (see for example [38]–[40]). Unlike concurrent works, our paper focuses explicitly on the system optimization under the fairness criteria in the multi-cell downlink with inter-cell cooperation. This particular angle allows us to illuminate aspects that are not present in other works, such as the distribution of the per-user throughput under fairness and, as a consequence, the design of the random scheduling scheme said before.

The remainder of this paper is organized as follows. In Section II, we describe the general finite-dimensional system model including the arbitrary clustering of cooperative BSs, formulate the system optimization problem, and provide its numerical solutions for a given channel realization. In Section III, we take the large system limit and present the large-system regime of the LZFB precoder and the optimization algorithm for user selection and power allocation. The opportunistic fairness scheduling scheme is also described in this section. The impact of non-perfect CSIT and training overhead is analyzed in Section IV. Numerical results and the low-complexity randomized scheduling scheme are presented in Section V and some concluding remarks are given in Section VI. The most lengthy and technical derivations are relegated into the appendices.

## II. FINITE DIMENSIONAL SYSTEM

### A. System Setup

Consider a cellular system formed by $M$ BSs, with $\gamma N$ antennas each, and $KN$ single-antenna user terminals spatially distributed in the coverage area. We assume that the users are divided into $K$ co-located "user groups" with $N$ users each. Users in the same group are statistically equivalent: they see the same pathloss coefficients from all BSs and their small-scale fading channel coefficients are i.i.d.. The received signal vector $\mathbf{y}_k = [y_{k,1} \cdots y_{k,N}]^\mathsf{T} \in \mathbb{C}^N$ for the $k$-th user group is given by

$$\mathbf{y}_k = \sum_{m=1}^{M} \alpha_{m,k} \mathbf{H}_{m,k}^\mathsf{H} \mathbf{x}_m + \mathbf{n}_k \tag{1}$$

where $\alpha_{m,k}$ and $\mathbf{H}_{m,k}$ denote the distance dependent pathloss coefficient and $\gamma N \times N$ small-scale channel fading matrix from the $m$-th BS to the $k$-th user group, respectively, $\mathbf{x}_m = [x_{m,1} \cdots x_{m,\gamma N}]^\mathsf{T} \in \mathbb{C}^{\gamma N}$ is the transmitted signal vector of the $m$-th BS, subject to the power constraint $\mathrm{tr}\left(\mathrm{Cov}(\mathbf{x}_m)\right) \leq P_m$, and $\mathbf{n}_k = [n_{k,1} \cdots n_{k,N}]^\mathsf{T} \in \mathbb{C}^N$ denotes the additive white Gaussian noise (AWGN) at the user receivers. The elements of $\mathbf{n}_k$ and of $\mathbf{H}_{m,k}$ are i.i.d. $\mathcal{CN}(0,1)$.

A cooperative cell arrangement with $L$ cooperation clusters is defined by the BS partition $\{\mathcal{M}_1, \cdots, \mathcal{M}_L\}$ of the BS set $\{1, \cdots, M\}$ and the corresponding user group partition $\{\mathcal{K}_1, \cdots, \mathcal{K}_L\}$ of the user group set $\{1, \cdots, K\}$. We assume that the BSs in each cluster $\mathcal{M}_\ell$ act as a single distributed multi-antenna transmitter with $\gamma |\mathcal{M}_\ell| N$ antennas, perfectly coordinated by a central cluster controller, and serve users in groups $k \in \mathcal{K}_\ell$. The clusters do not cooperate and treat ICI from other clusters as noise. Assuming that each BS operates at its maximum individual transmit power, the ICI plus noise power at any user terminal in group $k \in \mathcal{K}_\ell$ is given by

$$\sigma_k^2 = 1 + \sum_{m \notin \mathcal{M}_\ell} \alpha_{m,k}^2 P_m. \tag{2}$$

Each cluster seeks to maximize its own objective function defined by the fairness scheduling. It is easy to show that, under the above system assumptions, the selfish optimal strategy that operates at maximum per-BS power is a Nash equilibrium of the system. At this Nash equilibrium, the clusters are effectively decoupled since the effect that other clusters have on each cluster $\ell$ is captured by the ICI terms in (2) that do not depend on the actual BS transmit covariances $\mathrm{Cov}(\mathbf{x}_m)$.

From the viewpoint of cluster $\ell$, the system is equivalent to a single-cell MIMO downlink channel with a modified channel matrix and noise levels and a per-BS power constraint. Therefore, from now on we focus on a given reference cluster $\ell = 1$ and, without loss of generality, we indicate the user groups in the reference cluster as $k = 1, \ldots, A$, with $A = |\mathcal{K}_1|$, and the BSs in $\mathcal{M}_1$ as $m = 1, \ldots, B$ with $B = |\mathcal{M}_1|$. After a convenient re-normalization of the channel coefficients, we arrive at the equivalent channel model for the reference cluster given by

$$\mathbf{y} = \mathbf{H}^{\mathsf{H}} \mathbf{x} + \mathbf{z} \tag{3}$$

with $\mathbf{y} \in \mathbb{C}^{AN}$, $\mathbf{x} \in \mathbb{C}^{\gamma BN}$, $\mathbf{z} \sim \mathcal{CN}(\mathbf{0}, \mathbf{I}_{AN})$ and the channel matrix $\mathbf{H} \in \mathbb{C}^{\gamma BN \times AN}$ is given by

$$\mathbf{H} = \begin{bmatrix} \beta_{1,1} \mathbf{H}_{1,1} & \cdots & \beta_{1,A} \mathbf{H}_{1,A} \\ \vdots & & \vdots \\ \beta_{B,1} \mathbf{H}_{B,1} & \cdots & \beta_{B,A} \mathbf{H}_{B,A} \end{bmatrix}, \tag{4}$$

where we define $\beta_{m,k} = \alpha_{m,k}/\sigma_k$. The pathloss coefficients are fixed constant, that depend only on the geometry of the system, and the small-scale fading coefficients are assumed to evolve according to a block-fading process, that changes independently from slot to slot and remains constant over each slot. This is representative of a typical situation where the distance between BSs and users changes significantly over a time-scale of the order of the tens of seconds, while the small-scale fading decorrelates completely within a few milliseconds [41]. Here, a "slot" indicates a block of channel uses over which the small-scale coefficients can be considered constant. The slot length (in channel uses) is approximately equal to the product of the channel coherence time and the channel coherence bandwidth [41].

## B. Downlink Scheduling Optimization Problem

Each cluster controller operates according to a *downlink scheduling* scheme that allocates instantaneously the transmission resource (signal dimensions and transmit power) to the users. Following [22], the scheduling problem is formulated as the maximization of a suitable strictly increasing and concave *network utility function* $g(\cdot)$ over the region of achievable ergodic rates (throughput region), which is convex by time-sharing. For users in group $k$, we define the *mean group throughput* $\overline{R}_k$ as the the arithmetic mean of the individual user throughputs, i.e., $\overline{R}_k = \frac{1}{N} \sum_{i=1}^{N} \overline{R}_k^{(i)}$. By the symmetry of the system in the users belonging to the same group, it turns out that for any achievable throughput point with given individual user throughputs $\{\overline{R}_k^{(i)}\}$, there exists an achievable throughput point such that all users in group $k$ have throughput $\overline{R}_k$. In other words, the cumulative throughput of users in the same group can always be distributed uniformly over these users, without changing the sum throughput. We assume that the network utility function gives the same priority to statistically equivalent users. This is captured by restricting $g(\cdot)$ to be Schur-concave [42].[2] It follows that the network utility function is always maximized at a point for which $\overline{R}_k^{(i)} = \overline{R}_k$ for all users $i$ in group $k$. Therefore, letting $\overline{\mathbf{R}} = (\overline{R}_1, \cdots, \overline{R}_A)$ and redefining the function $g(\cdot)$ to have $A$ arguments, the fairness scheduling problem is formulated directly in terms of the user mean group throughputs, as

$$\text{maximize} \quad g(\overline{\mathbf{R}})$$
$$\text{subject to} \quad \overline{\mathbf{R}} \in \overline{\mathcal{R}}, \tag{5}$$

where $\overline{\mathcal{R}}$ denotes the system $A$-dimensional achievable group throughput region. In particular, this work considers LZFB downlink precoding. Hence, $\overline{\mathcal{R}}$ indicates the group throughput region achievable by LZFB for the channel model (3), under the assumption of operating at the Nash equilibrium said above. A scheduling policy achieving the optimum group throughput point $\overline{\mathbf{R}}^\star$ solution of (5) consists of a rule that, at each scheduling slot, maps the available channel information $\mathbf{H}$ into a set of scheduled users, rates and transmit powers, such that the resulting long-term time averaged group rates converge to $\overline{\mathbf{R}}^\star$.

As a first step towards the solution of (5), we focus on the weighted *instantaneous* sum-rate maximization problem:

$$\text{maximize} \quad \sum_{k=1}^{A} \sum_{i=1}^{N} W_k^{(i)} R_k^{(i)}$$
$$\text{subject to} \quad \mathbf{R} \in \mathcal{R}_{\text{lzfb}}(\mathbf{H}) \tag{6}$$

where $W_k^{(i)}$ denotes the rate weight for user $i$ in group $k$, and $\mathcal{R}_{\text{lzfb}}(\mathbf{H})$ is the achievable *instantaneous* rate region of LZFB for given channel matrix $\mathbf{H}$. By "instantaneous", we mean that this rate region depends on the given channel realization $\mathbf{H}$, in contrast with the throughput region $\overline{\mathcal{R}}$, that depends on the statistics of $\mathbf{H}$. Realistically, we assume that $A \geq \gamma B$ (i.e., the number of users in the cluster is larger than or equal to the total number of BS antennas in the cluster) and that all coefficients $\beta_{m,k}$ are strictly positive. Therefore, we have $\text{rank}(\mathbf{H}) = \gamma BN$ almost surely. In this case, LZFB cannot serve simultaneously all users in the cluster, and the scheduler must *select* a subset of users not larger than $\gamma BN$, to be served at each slot.

It should be noticed at this point that, for the sake of completeness, we consider the weighted *instantaneous* sum-rate maximization problem (6) in the most general case where the weights $W_k^{(i)}$ are distinct for each individual user. As a matter of fact, because of the system symmetry said before, in the large system limit we will be interested in the solution for $W_k^{(i)} = W_k$ (same weight, and therefore same priority, for all users $i$ in the same group $k$).

---

[2]The class of $\alpha$-fairness network utility functions introduced in [22], including max-min and proportional fairness, satisfy this condition.

The solution of (6) is generally difficult, since it requires a search over all user subsets of cardinality less or equal to $\gamma BN$. Well-known approaches (see [31], [32]) consider the selection of a user subset in some greedy fashion, by adding users to the *active user set* one by one, until the objective function in (6) cannot be improved further. Moreover, even for a fixed set of active users, the problem of optimal LZFB precoding subject to a per-BS power constraint is non-trivial and has been recently addressed in [43]–[45] through fairly involved numerical algorithms. Because of these difficulties, problem (6) has so far escaped a clean analytical solution and most studies resorted to extensive and costly Monte Carlo simulation.

In order to overcome the above difficulties, we make the following simplifying assumptions: 1) The scheduler picks a fraction $\mu_k$ of users in group $k$ by random selection inside the group. The selected users are referred to as the *active users* of group $k$. The active user set selection is statistically independent over different scheduling slots; 2) The LZFB precoder is obtained by normalizing the columns of the Moore-Penrose pseudo-inverse of the channel matrix, although this choice is not necessarily optimal under the per-BS power constraint [43].

Under these assumptions, we let $\boldsymbol{\mu} = (\mu_1, \ldots, \mu_A)$ denote the fractions of active users in groups $1, \ldots, A$, respectively. For given $\boldsymbol{\mu}$, the corresponding effective channel matrix is given by

$$\mathbf{H}_{\boldsymbol{\mu}} = \begin{bmatrix} \beta_{1,1}\mathbf{H}_{1,1}(\mu_1) & \cdots & \beta_{1,A}\mathbf{H}_{1,A}(\mu_A) \\ \vdots & & \vdots \\ \beta_{B,1}\mathbf{H}_{B,1}(\mu_1) & \cdots & \beta_{B,A}\mathbf{H}_{B,A}(\mu_A) \end{bmatrix}, \tag{7}$$

where the blocks $\mathbf{H}_{m,k}(\mu_k)$ is a $\gamma N \times \mu_k N$ dimensional submatrix of $\mathbf{H}_{m,k}$. The user fractions must satisfy $\mu_k \in [0, 1]$ for each $k = 1, \ldots, A$ and $\mu \overset{\Delta}{=} \mu_{1:A} \leq \gamma B$ where we introduce the notation

$$\mu_{1:k} = \sum_{j=1}^{k} \mu_j. \tag{8}$$

Hence, we have $\text{rank}(\mathbf{H}_{\boldsymbol{\mu}}) = \mu N$ almost surely.

With LZFB precoding, the transmitted signal $\mathbf{x}_{\boldsymbol{\mu}}$ in given by

$$\mathbf{x}_{\boldsymbol{\mu}} = \mathbf{V}_{\boldsymbol{\mu}} \mathbf{Q}^{1/2} \mathbf{u} \tag{9}$$

where $\mathbf{u} \in \mathbb{C}^{\mu N}$ contains the users' information-bearing code symbols, statistically independent with mean zero and variance 1, $\mathbf{V}_{\boldsymbol{\mu}}$ is the precoding matrix with unit-norm columns, and $\mathbf{Q}$ is a diagonal matrix which contains the user transmit powers on the diagonal. The precoding matrix $\mathbf{V}_{\boldsymbol{\mu}}$ is obtained as follows. The Moore-Penrose (right) pseudo-inverse of $\mathbf{H}_{\boldsymbol{\mu}}^{\mathsf{H}}$ is given by

$$\mathbf{H}_{\boldsymbol{\mu}}^{+} = \mathbf{H}_{\boldsymbol{\mu}}(\mathbf{H}_{\boldsymbol{\mu}}^{\mathsf{H}}\mathbf{H}_{\boldsymbol{\mu}})^{-1}. \tag{10}$$

Then, we let $\mathbf{V}_{\boldsymbol{\mu}} = \mathbf{H}_{\boldsymbol{\mu}}^{+}\boldsymbol{\Lambda}_{\boldsymbol{\mu}}^{1/2}$ where the column-normalizing diagonal matrix $\boldsymbol{\Lambda}_{\boldsymbol{\mu}}$ contains the reciprocal of the squared norm of the columns of $\mathbf{H}_{\boldsymbol{\mu}}^{+}$ on the diagonal. Letting $\Lambda_k^{(i)}(\boldsymbol{\mu})$ denote the diagonal element of $\boldsymbol{\Lambda}_{\boldsymbol{\mu}}$ in position $\mu_{1:k-1}N + i$, for $i = 1, \ldots, \mu_k N$, we have

$$\Lambda_k^{(i)}(\boldsymbol{\mu}) = \frac{1}{\left[\left(\mathbf{H}_{\boldsymbol{\mu}}^{\mathsf{H}}\mathbf{H}_{\boldsymbol{\mu}}\right)^{-1}\right]_k^{(i)}}, \tag{11}$$

where $\left[\left(\mathbf{H}_{\boldsymbol{\mu}}^{\mathsf{H}}\mathbf{H}_{\boldsymbol{\mu}}\right)^{-1}\right]_k^{(i)}$ denotes the element in the corresponding position $\mu_{1:k-1}N + i$ of the main diagonal of the matrix $\left(\mathbf{H}_{\boldsymbol{\mu}}^{\mathsf{H}}\mathbf{H}_{\boldsymbol{\mu}}\right)^{-1}$. Rewriting (3) with (7) and (9) and noticing that $\mathbf{H}_{\boldsymbol{\mu}}^{\mathsf{H}}\mathbf{V}_{\boldsymbol{\mu}} = \boldsymbol{\Lambda}_{\boldsymbol{\mu}}^{1/2}$, we arrive at the "parallel" channel model for all active users in the form

$$\mathbf{y}_{\boldsymbol{\mu}} = \boldsymbol{\Lambda}_{\boldsymbol{\mu}}^{1/2}\mathbf{Q}^{1/2}\mathbf{u} + \mathbf{z}_{\boldsymbol{\mu}}. \tag{12}$$

The optimization of (6) for the channel model (12) is still involved, since the channel coefficients $\Lambda_k^{(i)}(\boldsymbol{\mu})$ depend on the active user fractions $\boldsymbol{\mu}$ in a complicated and non-convex way. As an intermediate step, we consider the solution of (6) for fixed user fractions $\boldsymbol{\mu}$.

### C. Power Allocation under Sum-power or Per-BS Power Constraints

We divide all channel matrix coefficients by $\sqrt{N}$ and multiply the BS input power constraints $P_m$ by $N$, thus obtaining an equivalent system where the channel coefficients have variance that scales as $1/N$. This is useful when considering the large-system limit for $N \to \infty$ in Section III.

Let $q_k^{(i)}$ denote the diagonal element in position $\mu_{1:k-1}N + i$ of $\mathbf{Q}$, corresponding to the power allocated to the $i$-th user of group $k$. The sum-power constraint is given by

$$\frac{1}{N}\text{tr}(\mathbf{Q}) = \frac{1}{N}\sum_{k=1}^{A}\sum_{i=1}^{\mu_k N}q_k^{(i)} \leq P_{\text{sum}} \tag{13}$$

where $P_{\text{sum}} = \sum_{m=1}^{B}P_m$. In order to express the per-BS power constraint, let $\boldsymbol{\Phi}_m$ denote a diagonal matrix with all zeros, but for $\gamma N$ consecutive "1", corresponding to positions from $(m-1)\gamma N + 1$ to $m\gamma N$ on the main diagonal. Then, the per-BS power constraint is expressed in terms of the partial trace of the transmitted signal covariance matrix as

$$\frac{1}{N}\text{tr}\left(\boldsymbol{\Phi}_m\mathbf{V}_{\boldsymbol{\mu}}\mathbf{Q}\mathbf{V}_{\boldsymbol{\mu}}^{\mathsf{H}}\right) \leq P_m, \quad m = 1, \ldots, B \tag{14}$$

More explicitly, (14) can be written in terms of the powers $q_k^{(i)}$ as

$$\sum_{k=1}^{A} \sum_{i=1}^{\mu_k N} q_k^{(i)} \theta_{m,k}^{(i)}(\boldsymbol{\mu}) \leq P_m, \quad m = 1, \ldots, B \tag{15}$$

where we define the coefficients

$$\theta_{m,k}^{(i)}(\boldsymbol{\mu}) = \frac{1}{N} \sum_{\ell=(m-1)\gamma N+1}^{m\gamma N} \left| \left[ \mathbf{V}_{\boldsymbol{\mu}} \right]_{\ell,k}^{(i)} \right|^2 \tag{16}$$

and where $\left[ \mathbf{V}_{\boldsymbol{\mu}} \right]_{\ell,k}^{(i)}$ denotes the element of $\mathbf{V}_{\boldsymbol{\mu}}$ corresponding to the $\ell$-th row and the $(\mu_{1:k-1}N + i)$-th column. Since $\mathbf{V}_{\boldsymbol{\mu}}$ has unit-norm columns, then $\sum_{m=1}^{B} \theta_{m,k}^{(i)}(\boldsymbol{\mu}) = 1/N$ for all $k, i$.

For fixed user fractions $\boldsymbol{\mu}$, the weighted instantaneous sum-rate maximization (6) reduces to

$$\text{maximize} \quad \sum_{k=1}^{A} \sum_{i=1}^{\mu_k N} W_k^{(i)} \log(1 + \Lambda_k^{(i)}(\boldsymbol{\mu}) q_k^{(i)}) \tag{17}$$

subject to (13) in the case of sum-power constraint, or to (15) for the case of per-BS power constraint.

The solution of (17) subject to the sum-power constraint is immediately given by the water-filling formula

$$q_k^{(i)} = \left[ \frac{W_k^{(i)}}{\lambda} - \frac{1}{\Lambda_k^{(i)}(\boldsymbol{\mu})} \right]_+ \tag{18}$$

where $\lambda \geq 0$ is the Lagrange multiplier corresponding to the sum-power constraint.

In the case of per-BS power constraint, we can use Lagrange duality and the sub-gradient iteration method as given in the following. The Lagrangian function for (17) is given by

$$\mathcal{L}(\mathbf{q}, \boldsymbol{\lambda}) = \sum_{k=1}^{A} \sum_{i=1}^{\mu_k N} W_k^{(i)} \log(1 + \Lambda_k^{(i)}(\boldsymbol{\mu}) q_k^{(i)}) - \boldsymbol{\lambda}^{\mathsf{T}} \left[ \boldsymbol{\Theta} \mathbf{q} - \mathbf{P} \right] \tag{19}$$

where $\boldsymbol{\lambda} \geq 0$ is a vector of dual variables corresponding to the $B$ BS power constraints, $\boldsymbol{\Theta}$ is the $B \times \mu N$ matrix containing the coefficients $\theta_{m,k}^{(i)}(\boldsymbol{\mu})$ and $\mathbf{P} = (P_1, \ldots, P_B)^{\mathsf{T}}$. The KKT conditions are given by

$$\frac{\partial \mathcal{L}}{\partial q_k^{(i)}} = W_k^{(i)} \frac{\Lambda_k^{(i)}(\boldsymbol{\mu})}{1 + \Lambda_k^{(i)}(\boldsymbol{\mu}) q_k^{(i)}} - \boldsymbol{\lambda}^{\mathsf{T}} \boldsymbol{\theta}_k^{(i)} \leq 0 \tag{20}$$

where $\boldsymbol{\theta}_k^{(i)}$ indicates the column of $\boldsymbol{\Theta}$ containing the coefficients $\theta_{m,k}^{(i)}(\boldsymbol{\mu})$ for $m = 1, \ldots, B$. Solving for $q_k^{(i)}$, we find

$$q_k^{(i)}(\boldsymbol{\lambda}) = \left[ \frac{W_k^{(i)}}{\boldsymbol{\lambda}^{\mathsf{T}} \boldsymbol{\theta}_k^{(i)}} - \frac{1}{\Lambda_k^{(i)}(\boldsymbol{\mu})} \right]_+ \tag{21}$$

Replacing this solution into $\mathcal{L}(\mathbf{q}, \boldsymbol{\lambda})$, we solve the dual problem by minimizing $\mathcal{L}(\mathbf{q}(\boldsymbol{\lambda}), \boldsymbol{\lambda})$ with respect to $\boldsymbol{\lambda} \geq 0$. It is immediate to check that for any $\boldsymbol{\lambda}' \geq 0$,

$$\mathcal{L}(\mathbf{q}(\boldsymbol{\lambda}'), \boldsymbol{\lambda}') \geq \mathcal{L}(\mathbf{q}(\boldsymbol{\lambda}), \boldsymbol{\lambda}')$$

$$= (\boldsymbol{\lambda}' - \boldsymbol{\lambda})^{\mathsf{T}} (\mathbf{P} - \boldsymbol{\Theta} \mathbf{q}(\boldsymbol{\lambda})) + \mathcal{L}(\mathbf{q}(\boldsymbol{\lambda}), \boldsymbol{\lambda}) \tag{22}$$

Therefore, $(\mathbf{P} - \boldsymbol{\Theta}\mathbf{q}(\boldsymbol{\lambda}))$ is a subgradient for $\mathcal{L}(\mathbf{q}(\boldsymbol{\lambda}), \boldsymbol{\nu})$. It follows that the dual problem can be solved by a simple $B$-dimensional subgradient iteration over the vector of dual variables $\boldsymbol{\lambda}$.

## III. Large System Limit

In this section, we consider the limit of the above instantaneous rate maximization problems for $N \to \infty$ and fixed $\gamma, A, B$, and $\boldsymbol{\mu}$. We shall see in Section III-D that the weights in the weighted instantaneous sum-rate maximization are recursively calculated by the scheduler that solves the general network utility maximization problem (5). For Schur-concave $g(\cdot)$ and in the large system limit, where the "instantaneous" channel gains $\Lambda_k^{(i)}(\boldsymbol{\mu})$ freeze to deterministic limits that depend only on the group index $k$ and not on the individual user index $i$ (see Theorem 1 below), these weights must be identical for all users in the same group. Since the weighted-sum rate maximization problem is used here as an intermediate step to devise the scheduling rule that solves (5), from now on we restrict to the case $W_k^{(i)} = W_k$ for all users $i$ in group $k$.

### A. Asymptotic Analysis

We start by finding the large system limit expression for the coefficients $\Lambda_k^{(i)}(\boldsymbol{\mu})$. This is provided by:

*Theorem 1:* For all $i = 1, \ldots, \mu_k N$, the following limit holds almost surely:

$$\lim_{N \to \infty} \Lambda_k^{(i)}(\boldsymbol{\mu}) = \Lambda_k(\boldsymbol{\mu}) = \gamma \sum_{m=1}^{B} \beta_{m,k}^2 \eta_m(\boldsymbol{\mu}) \tag{23}$$

where $(\eta_1(\boldsymbol{\mu}), \ldots, \eta_B(\boldsymbol{\mu}))$ is the unique solution in $[0,1]^B$ of the system of fixed-point equations

$$\eta_m = 1 - \sum_{q=1}^{A} \mu_q \frac{\eta_m \beta_{m,q}^2}{\gamma \sum_{\ell=1}^{B} \eta_\ell \beta_{\ell,q}^2}, \quad m = 1, \ldots, B, \tag{24}$$

with respect to the variables $\boldsymbol{\eta} = \{\eta_m\}$.

*Proof:* See Appendix A. ∎

As anticipated above, the limit (23) depends only on $k$ (user group index) and not on $i$ (user index within the group), consistently with the fact that, in our model, users in the same co-located group are statistically equivalent. An immediate consequence is that in the limit for $N \to \infty$, for the case $W_k^{(i)} = W_k$ considered here, the waterfilling equation (18) yields equal power allocation $q_k^{(i)} = q_k$ for all active users $i$ in group $k$.

Next, we consider the per-BS power constraint given in (15). By the system symmetry and for the sake of analytical tractability, also in this case we restrict to uniform power allocation $q_k^{(i)} = q_k$ for all active users $i$ in the same group $k$. Replacing this in the constraint (15), we obtain

$$\sum_{k=1}^{A} q_k \theta_{m,k}(\boldsymbol{\mu}) \leq P_m, \quad m = 1, \ldots, B, \tag{25}$$

where we define

$$\theta_{m,k}(\boldsymbol{\mu}) = \sum_{i=1}^{\mu_k N} \theta_{m,k}^{(i)}(\boldsymbol{\mu}) = \frac{1}{N} \sum_{i=1}^{\mu_k N} \sum_{\ell=1+(m-1)\gamma N}^{m\gamma N} \left| \left[ \mathbf{V}_{\boldsymbol{\mu}} \right]_{\ell,k}^{(i)} \right|^2. \tag{26}$$

It is interesting to notice that $\theta_{m,k}(\boldsymbol{\mu})$ is the squared Frobenius norm (normalized by $N$) of the submatrix of $\mathbf{V}_{\boldsymbol{\mu}}$ corresponding to the users in group $k$ (columns from $\mu_{1:k-1}N + 1$ to $\mu_{1:k}N$) and the antennas of BS $m$ (rows from $(m-1)\gamma N + 1$ to $m\gamma N$).

We hasten to point out that the choice $q_k^{(i)} = q_k$ does not follow immediately from the KKT conditions (20), even assuming $W_k = W_k^{(i)}$ and in the limit of $\Lambda_k^{(i)}(\boldsymbol{\mu}) \to \Lambda_k(\boldsymbol{\mu})$, since the terms $\theta_{m,k}^{(i)}(\boldsymbol{\mu})$ may depend on $i$ and not only on $k$ even in the large system limit. As a matter of fact, Theorem 2 below yields the convergence of $\theta_{m,k}(\boldsymbol{\mu}) = \sum_{i=1}^{\mu_k N} \theta_k^{(i)}(\boldsymbol{\mu})$ to a deterministic limit. Notice that each individual term $\theta_{m,k}^{(i)}(\boldsymbol{\mu})$ vanishes as $N \to \infty$ since, by construction, $\sum_{m=1}^{B} \theta_{m,k}^{(i)}(\boldsymbol{\mu}) = 1/N$. Based on numerical evidence, we conjecture that $|N\theta_{m,k}^{(i)}(\boldsymbol{\mu}) - \theta_{m,k}(\boldsymbol{\mu})| \to 0$ as $N \to \infty$, for all $i$ in group $k$. However, proving the convergence of the individual random variables $N\theta_{m,k}^{(i)}(\boldsymbol{\mu})$ to the same deterministic limit independent of $i$ has resisted our efforts. In conclusions, beyond the sake of analytical tractability and system symmetry considerations, we also conjecture that the symmetric power allocation $q_k^{(i)} = q_k$ for users in the same group is also optimal for the case of per-BS power constraint, in the limit of $N \to \infty$. The next result yields an analytical expression for the large-system limits of the coefficients $\theta_{m,k}(\boldsymbol{\mu})$:

*Theorem 2:* For all $m, k$, the following limit holds almost surely:

$$\lim_{N \to \infty} \theta_{m,k}(\boldsymbol{\mu}) = \frac{\mu_k \eta_m^2(\boldsymbol{\mu}) \left( \beta_{m,k}^2 + \xi_{m,k} \right)}{\sum_{\ell=1}^{B} \eta_\ell(\boldsymbol{\mu}) \beta_{\ell,k}^2} \tag{27}$$

where $\boldsymbol{\xi}_m = (\xi_{m,1}, \ldots, \xi_{m,A})^\mathsf{T}$ is the solution to the linear system

$$\left[ \mathbf{I} - \gamma \mathbf{M} \right] \boldsymbol{\xi}_m = \gamma \mathbf{M} \mathbf{b}_m \tag{28}$$

where $\mathbf{M}$ is the $A \times A$ matrix

$$\mathbf{M} = \left[ \sum_{\ell=1}^{B} \eta_\ell^2(\boldsymbol{\mu}) \mathbf{b}_\ell \mathbf{b}_\ell^\mathsf{T} \right] \operatorname{diag}\left( \frac{\mu_1}{\Lambda_1^2(\boldsymbol{\mu})}, \ldots, \frac{\mu_A}{\Lambda_A^2(\boldsymbol{\mu})} \right) \tag{29}$$

and $\mathbf{b}_\ell = (\beta_{\ell,1}^2, \ldots, \beta_{\ell,A}^2)^\mathsf{T}$, and the coefficients $\{\eta_m(\boldsymbol{\mu})\}$ and $\{\Lambda_k(\boldsymbol{\mu})\}$ are provided by Theorem 1.

*Proof:* See Appendix B. ∎

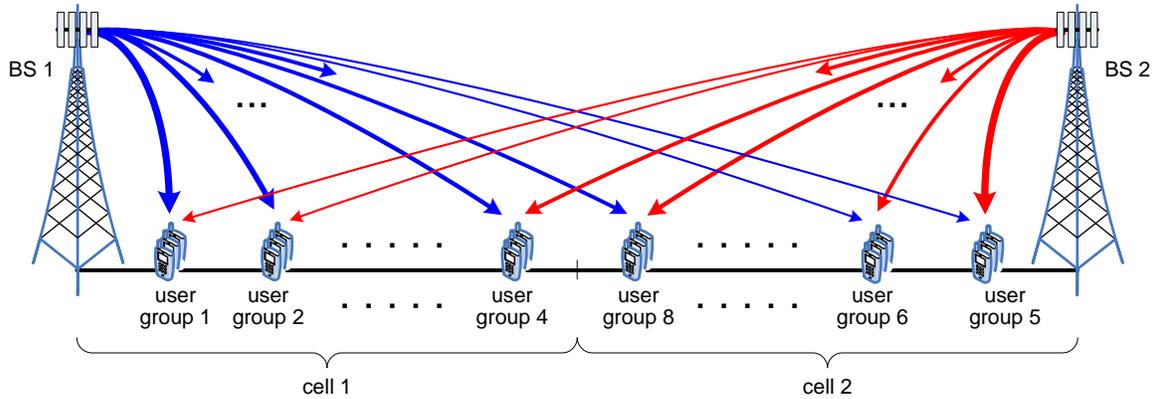

Fig. 1.  Linear one-sided 2-cell model with $B = 2$ BSs and $A = 8$ user groups.

*1) Simplifications for Symmetric System:* Under some special symmetry conditions, the general problem can be significantly simplified. In particular, we assume that $B$ divides $A$, let $A' = A/B$, and that the $B \times A$ matrix of channel gains $\boldsymbol{\beta} = [\beta^2_{m,k}]$ can be partitioned into $A'$ circulant submatrices of size $B \times B$, i.e., such that each submatrix has the property that all rows are cyclic shifts of the first row and all columns are cyclic shifts of the first column. We shall refer to these submatrices as "circulant blocks." Also, we define users' *equivalence classes* as the sets of user groups whose corresponding columns of $\boldsymbol{\beta}$ form circulant blocks. Then, the set of $A$ user groups is partitioned into $A'$ equivalence classes. We re-index the user groups such that groups $\{(j-1)A' + i : j = 1, \ldots, B\}$ form the $i$-th equivalence class, for $i = 1, \ldots, A'$. To fix ideas, consider Fig. 1 showing a one-dimensional cellular system comprising 2 BSs and and 8 symmetrically located user groups. Since the channel gain coefficients depend on distance, because of the symmetric layout, the matrix $\boldsymbol{\beta}$ is given by

$$\boldsymbol{\beta} = \begin{bmatrix} a & b & c & c & f & e & e & d \\ f & e & e & d & a & b & c & c \end{bmatrix} \tag{30}$$

for some positive numbers $a, b, c, d, e, f$. We notice that this matrix can be decomposed into the $A' = 4$ circulant blocks

$$\begin{bmatrix} a & f \\ f & a \end{bmatrix}, \quad \begin{bmatrix} b & e \\ e & b \end{bmatrix}, \quad \begin{bmatrix} c & e \\ e & c \end{bmatrix}, \quad \begin{bmatrix} c & d \\ d & c \end{bmatrix}$$

When this symmetry condition holds, users in the same equivalence class are statistically equivalent, up to renumbering of the BSs. This is because such groups (e.g., user group pairs $(1, 5)$, $(2, 6)$, $(3, 7)$, and $(4, 8)$ in the example) are equivalent as far as the "landscape" of channel gain coefficients seen collectively by the cluster BSs.

Motivated again by the Schur-concavity of the network utility function and by the symmetry of all users in the same equivalent class, for the sake of finding an analytically tractable solution and an overall problem simplification, we shall restrict to the resource allocation that gives to all groups $k$ in the same equivalence class $i$ the same transmit power and active user fraction (symmetric solution). We indicate these common powers and active user fractions by $q_i'$ and $\mu_i'$, respectively, for $i = 1, \ldots, A'$, such that $q_{(j-1)A'+i} = q_i'$ and $\mu_{(j-1)A'+i} = \mu_i'$ for all $j = 1, \ldots, B$. In this case, for any $m$, we have

$$\sum_{q=1}^{A} \mu_q \frac{\beta_{m,q}^2}{\gamma \sum_{\ell=1}^{B} \beta_{\ell,q}^2} = \frac{1}{\gamma} \sum_{i=1}^{A'} \mu_i' \sum_{j=1}^{B} \frac{\beta_{m,(j-1)A'+i}^2}{\sum_{\ell=1}^{B} \beta_{\ell,(j-1)A'+i}^2} = \frac{1}{\gamma} \sum_{i=1}^{A'} \mu_i' = \frac{\mu}{\gamma B}$$

where we used the fact that, by the symmetry condition, $\sum_{j=1}^{B} \frac{\beta_{m,(j-1)A'+i}^2}{\sum_{\ell=1}^{B} \beta_{\ell,(j-1)A'+i}^2} = 1$ and $\sum_{i=1}^{A'} \mu_i' = \frac{1}{B} \sum_{q=1}^{A} \mu_q = \frac{\mu}{B}$. It follows that the solution of the fixed point equation (24) is given explicitly by

$$\eta_m(\boldsymbol{\mu}) = 1 - \frac{\mu}{\gamma B} \tag{31}$$

which is independent of $m$, and (23) yields

$$\Lambda_k(\boldsymbol{\mu}) = \gamma \left(1 - \frac{\mu}{\gamma B}\right) \sum_{m=1}^{B} \beta_{m,k}^2. \tag{32}$$

Notice that for all groups $k$ in class $i$, i.e., for all $k = (j-1)A'+i$, $\forall j = 1, \ldots, B$, the sum $\sum_{m=1}^{B} \beta_{m,k}^2$ is a constant independent of $k$. Therefore, as expected by the symmetry condition, all active users in the same equivalence class have the same LZFB channel gains. With some abuse of notation, we define $\beta_i^2 = \sum_{m=1}^{B} \beta_{m,k}^2$ for all groups $k$ in the $i$-th equivalence class. In addition, we have:

*Lemma 1:* For symmetric systems with $\mu_k = \mu_i'$ for all $k$ in equivalence class $i$, we have

$$\theta_{m \oplus_B j, k}(\boldsymbol{\mu}) = \theta_{m, k \oplus_A A' j}(\boldsymbol{\mu}), \tag{33}$$

where we define the cyclic index shifts $m \oplus_B j \triangleq ((m+j-1) \mod B) + 1$ and $k \oplus_A A'j \triangleq ((k + A'j - 1) \mod A) + 1$, and furthermore, if $q_k = q_i'$ for all $k$ in equivalence class $i$, the transmit power of BS $m$ is independent of $m$, i.e.,

$$\sum_{k=1}^{A} q_k \theta_{m,k}(\boldsymbol{\mu}) = \sum_{i=1}^{A'} q_i' \mu_i'. \tag{34}$$

*Proof:* See Appendix C. ∎

As an immediate corollary, we have that if all the BSs in the cluster have the equal power constraint, i.e., $P_1 = \ldots = P_B = P$, then the per-BS power constraint (25) coincides with the sum power constraint with $P_{\text{sum}} = BP$.

| m \ k | 1 | 2 | 3 | 4 | 5 | 6 | 7 | 8 |
|---|---|---|---|---|---|---|---|---|
| 1 | 0.325 | 0.311 | 0.454 | 0.565 | 0.175 | 0.189 | 0.296 | 0.435 |
| 2 | 0.175 | 0.189 | 0.296 | 0.435 | 0.325 | 0.311 | 0.454 | 0.565 |

*Example 1:* Consider the two cell symmetric model in Fig. 1 and assume that the two BSs are cooperating ($B$=2) and serving $A$=8 user groups, the channel gain coefficients are given as (30) with $[a\ b\ c\ d\ e\ f] = [1.5\ 1.3\ 1.0\ 0.5\ 0.3\ 0.2]$, and the antenna ratio is given as $\gamma = 4$. If $\boldsymbol{\mu} = [0.5\ 0.5\ 0.75\ 1\ 0.5\ 0.5\ 0.75\ 1]$, the asymptotic values of $\theta_{m,k}(\boldsymbol{\mu})$ is given as in Table I. We notice that the same block-circulant form of the gain matrix $\boldsymbol{\beta}$ appears in the matrix $\{\theta_{m,k}(\boldsymbol{\mu})\}$, as given by Lemma 1. ∎

### B. Weighted Sum-rate Maximization

Using the asymptotic results obtained before, we consider the weighted sum-rate maximization problem in (17). First, we focus on the sum-power constraint (13). As noticed before, in the large system limit and in the case of uniform weights $W_k^{(i)} = W_k$ over each group $k$, the weighted sum-rate maximization solution yields that all the active users in the same group are allocated the same power and therefore achieve the same instantaneous rate. In these conditions, from (23) we have that the instantaneous mean group rate converges to $\frac{1}{N} \sum_{i=1}^{\mu_k N} R_k^{(i)} \rightarrow \mu_k R_k$ with

$$R_k = \log\left(1 + \Lambda_k(\boldsymbol{\mu})q_k\right) \tag{35}$$

Notice that $R_k$ in the large-system limit is a deterministic quantity, therefore, the mean group throughput in this regime is also given by $\overline{R}_k = \mu_k R_k$.

Using (24), we can write the large-system limit weighted sum-rate maximization problem subject to the sum-power constraint in the form:

$$\text{maximize} \quad \sum_{k=1}^{A} W_k \mu_k \log \left( 1 + \gamma \left( \sum_{m=1}^{B} \beta_{m,k}^2 \eta_m \right) q_k \right) \tag{36a}$$

$$\text{subject to} \quad \sum_{k=1}^{A} \mu_k q_k \leq P_{\text{sum}}, \quad \sum_{k=1}^{A} \mu_k \leq \gamma B, \tag{36b}$$

$$\eta_m = 1 - \sum_{k=1}^{A} \mu_k \frac{\eta_m \beta_{m,k}^2}{\gamma \sum_{\ell=1}^{B} \eta_\ell \beta_{\ell,k}^2}, \quad m = 1, \ldots, B \tag{36c}$$

$$0 \leq \eta_m \leq 1, \quad m = 1, \ldots, B \tag{36d}$$

$$q_k \geq 0, \quad 0 \leq \mu_k \leq 1, \quad k = 1, \ldots, A \tag{36e}$$

This problem is generally non-convex in $\mathbf{q}$, $\boldsymbol{\mu}$ and $\boldsymbol{\eta}$. However, for fixed $\boldsymbol{\eta}$ and $\boldsymbol{\mu}$, it is convex in $\mathbf{q}$, and the solution is given by water-filling (see also (18)):

$$q_k = \left[ \frac{W_k}{\lambda} - \frac{1}{\gamma \left( \sum_{m=1}^{B} \beta_{m,k}^2 \eta_m \right)} \right]_+ \tag{37}$$

For fixed $\boldsymbol{\eta}$ and $\mathbf{q}$, we have a linear program with respect to $\boldsymbol{\mu}$. Finally, for fixed $\boldsymbol{\mu}$ and $\mathbf{q}$ the problem is degenerate with respect to $\boldsymbol{\eta}$ since the equality constraint (36c), that corresponds to the fixed-point equation (24), has a unique solution $\boldsymbol{\eta} \in [0, 1]^B$ for all feasible $\boldsymbol{\mu}$.

In the symmetric system case with the conditions given in the previous section, we have that user groups in the same equivalence class are completely symmetric, since the limits $\Lambda_k(\boldsymbol{\mu})$ depend only on the equivalence class and not on the specific user group in the class. Then, the optimization problem in the symmetric case reduces to optimizing the powers $q_i'$ and the fractions $\mu_i'$ for the equivalence classes $i = 1, \ldots, A'$. Letting $\mu' = \sum_{i=1}^{A'} \mu_i' = \mu/B$, we can state the optimization problem in the symmetric case as:

$$\text{maximize} \quad B \sum_{i=1}^{A'} W_i \mu_i' \log \left( 1 + \gamma \left( 1 - \frac{\mu'}{\gamma} \right) \beta_i^2 q_i' \right) \tag{38a}$$

$$\text{subject to} \quad B \sum_{i=1}^{A'} \mu_i' q_i' \leq P_{\text{sum}}, \tag{38b}$$

$$\sum_{i=1}^{A'} \mu_i' \leq \gamma, \tag{38c}$$

$$q_i' \geq 0, \quad 0 \leq \mu_i' \leq 1, \quad i = 1, \ldots, A' \tag{38d}$$

The net effect of the symmetry is a sort of "resource pooling": the system with a cluster of $B$ cooperating BSs reduces to an equivalent single-BS system with total transmit power $P_{\text{sum}}/B$, load $\mu' = \mu/B$, $A' = A/B$ user classes, and equivalent channel path gains $\beta_i^2 = \sum_{m=1}^{B} \beta_{m,k}^2$ given by the combination of the path gains from all BSs in the cluster to the user groups in the $i$-th equivalence class.

As far the per-BS power constraint is concerned, the power constraint in problem (36) must be replaced by (25) where the coefficients $\{\theta_{m,k}(\boldsymbol{\mu})\}$ are provided by Theorem 2. Finally, in the symmetric case, using Lemma 1 and assuming $P_m = P$ for all $m$, the per-BS power constraint reduces to

$$\sum_{i=1}^{A'} q_i' \mu_i' \leq P, \quad \text{for all} \ \ m = 1, \ldots, B.$$

Notice that the above set of constraints is identical for all BSs, and by summing over $m$ we obtain the equivalent constraint $B \sum_{i=1}^{A'} q_i' \mu_i' \leq BP = P_{\text{sum}}$, which coincides with the sum power constraint in (38), as anticipated before.

## C. Optimization of the User Fractions and Powers

While (36) is still a non-convex problem in $(\mathbf{q}, \boldsymbol{\mu})$, we can find near-optimal solutions by borrowing from the greedy user selection heuristic used in the finite-dimensional case (see [31], [32]). In particular, we consider the approach of incrementing user fractions $\mu$ sequentially in very small steps, $\Delta\mu \ll 1$, until the objective function value cannot be increased any longer. If we take the infinitesimal of $\Delta\mu$, this is equivalent to greedy user selection in the large system limit where $\Delta\mu$ denoting the fraction of one user to the total number of users goes to zero. We start from $\boldsymbol{\mu} = \mathbf{0}$ and at each step we find $k$ such that incrementing $\mu_k$ by $\Delta\mu$ yields the largest improvement and the resulting new $\boldsymbol{\mu}$ is feasible. For the tentative configuration of the fractions $\boldsymbol{\mu}$, the corresponding power allocation is obtained from the waterfilling solution. We stop when no further increment can improve the objective function value. The detailed description is given in the following:

1) Initialize variables such that $n = 0$, $R_{\text{wsr}}(0) = 0$, $\boldsymbol{\mu} = \mathbf{0}$, and $\mu = 0$.

2) Set $n \leftarrow n + 1$. For $\Delta\mu \ll 1$, set $\boldsymbol{\mu}^{(k)} = \boldsymbol{\mu} + \Delta\mu \mathbf{e}_k$ (note: $\mathbf{e}_k$ denotes a vector of length $A$ of all zeros with a single 1 in position $k$), for $k \in \mathcal{S} = \{j : \mu_j + \Delta\mu \leq 1, \forall j\}$. If $\mathcal{S}$ is empty or $\mu + \Delta\mu > \gamma$, then exit and keep the current $\boldsymbol{\mu}$ and the corresponding rates as the final values of the algorithm. Otherwise, compute the tentative weighted sum rate value $R_{\text{wsr}}^{(k)}$ for each $k$, by solving the optimization problem in (36) for fixed $\boldsymbol{\mu}^{(k)}$ with the waterfilling power allocation.

3) Let $\widehat{k} = \arg\max_{k \in \mathcal{S}} R_{\text{wsr}}^{(k)}$ and set $R_{\text{wsr}}(n) = R_{\text{wsr}}^{(\widehat{k})}$.

4) If $R_{\text{wsr}}(n) > R_{\text{wsr}}(n-1)$, then set $\boldsymbol{\mu} \leftarrow \boldsymbol{\mu}^{(\widehat{k})}$, $\mu \leftarrow \mu + \Delta\mu$ and go back to step 2.

5) Otherwise, if $R_{\text{wsr}}(n) \leq R_{\text{wsr}}(n-1)$, exit and take the current $\boldsymbol{\mu}$ and the corresponding rates as the final values of the algorithm.

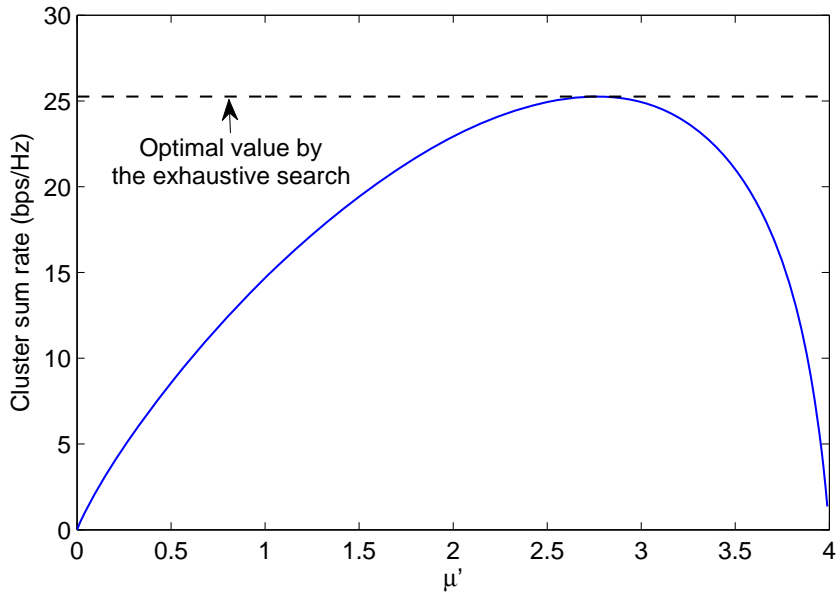

Fig. 2. Cluster ($B = 2$) sum rate of the proposed optimization algorithm as $\mu'$ increases from 0 to $\gamma = 4$.

Fig. 2 shows the sum rate versus $\mu' = \mu/B$ for the symmetric system of Example 1, with $P = 15$ dB and $W_i = 1$, $\forall i$. The optimization with respect to $\boldsymbol{\mu}'$ and $\mathbf{q}'$ is obtained by applying the algorithm described above, and the value of the objective function (sum rate) is compared with the globally optimal value obtained from exhaustive search. The exhaustive search finds the optimal weighted sum rate by searching over $\boldsymbol{\mu}' \in [0,1]^{A'}$, subject to $\sum_{i=1}^{A'} \mu'_i \leq \gamma$. If we discretize this domain with $\Delta\mu$ step size for each dimension, the computational complexity of the exhaustive algorithm is $O((1/\Delta\mu)^{A'})$, whereas the proposed greedy algorithm has complexity $O(A'\gamma/\Delta\mu)$. In order to obtain the curve in Fig. 2 we removed the comparison between $R_{\mathrm{wsr}}(n)$ and $R_{\mathrm{wsr}}(n-1)$ in step 4, so that the algorithm does not stop when it cannot improve the objective function value any longer, and returns a value of the objective function over the whole range $\mu' \in [0,\gamma]$. When $\Delta\mu = 0.01$, the greedy algorithm achieves the same optimal value of the exhaustive search, at $\mu' = 2.76$.

### D. Network Utility Function Maximization

In general, the solution of (6) (or (36) in the large system limit) for the case $A > \gamma B$ (more users than antennas) yields an unbalanced distribution of instantaneous rates, where some user classes are not served at all (we have $\mu_k = 0$ for some $k$). Hence, for a general strictly concave network utility function $g(\cdot)$, the ergodic rate region $\overline{\mathcal{R}}$ requires time-sharing even in the asymptotic large-system case. Therefore, the network utility maximization problem (5) is not generally amenable to a closed-form solution. However, the solution can be computed to any level of accuracy by using a method inspired by the stochastic optimization approach of [20]. Interestingly, the same algorithm can be used both for the computation of the optimum throughput point in the large system limit, and for the actual downlink scheduling algorithm, applied on a slot-by-slot basis to the actual finite-dimensional system. In the former case, the algorithm is equivalent to Lagrangian iteration where the "virtual queues" (to be defined in the following) plays the role of Lagrange multipliers. In the latter case, when applied to the finite dimensional system, the algorithm performs a stochastic "Lyapunov drift" optimization (see [20]).

For each user group $k = 1, \ldots, A$, we define a *virtual queue* that evolves according to

$$Q_k(t+1) = [Q_k(t) - r_k(t)]_+ + a_k(t) \tag{39}$$

where $r_k(t)$ denotes the virtual service rate and $a_k(t)$ the virtual arrival process. The queues are initialized by $Q_k(0) = r_k(0) = 0$. At each algorithm iteration $t = 1, 2, \ldots$, the virtual arrival processes is given by $a_k(t) = a_k^\star$, where $\mathbf{a}^\star$ is the solution of the convex programming problem:

$$\text{maximize} \quad V g(\mathbf{a}) - \sum_{k=1}^{A} Q_k(t) a_k$$

$$\text{subject to} \quad 0 \le a_k \le a_{\max}, \quad \forall \, k \tag{40}$$

and where $V, a_{\max} > 0$ are suitably chosen constant parameters that determine the convergence properties of the algorithm. The service rates are given by

$$r_k(t) = \mu_k(t) \log \left( 1 + \gamma \left( \sum_{m=1}^{B} \beta_{m,k}^2 \eta_m(t) \right) q_k(t) \right)$$

where $(\boldsymbol{\mu}(t), \mathbf{q}(t), \boldsymbol{\eta}(t))$ is the solution of (36) for weights $W_k = Q_k(t)$. Then, the virtual queues are updated according to (39). The theory developed in [20] (see also [21]) ensures the following result. Let $\mathbf{r}(t)$ denote the vector of service rates generated by the above iterative algorithm. Then,

$$\liminf_{t \to \infty} g \left( \frac{1}{t} \sum_{\tau=0}^{t-1} \mathbf{r}(\tau) \right) \ge g(\overline{\mathbf{R}}^\star) - \frac{\mathcal{K}}{V} \tag{41}$$

where $\overline{\mathbf{R}}^\star$ is the solution of (5) and $\mathcal{K}$ is a constant that depends on the system parameters and on $a_{\max}$. In particular, using the results in [21] we can show the bound

$$\mathcal{K} \le \frac{A}{2} \left( a_{\max}^2 + \log^2 \left( 1 + \gamma \max\{ \beta_{m,k}^2 : \forall \, m, k \} P_{\text{sum}} \right) \right)$$

By choosing $V$ and $a_{\max}$ appropriately, a desired tradeoff between the accuracy of the approximation of the optimum point and the convergence speed of the iterative algorithm can be ensured.

It should be noticed that if we use the greedy optimization of the user fractions as described in Section III-C instead of the exact solution of (38), then the performance guarantee (41) is no longer valid. However, the algorithm ensures that the throughput point that maximizes $g(\cdot)$ over the ergodic rate region achievable with the (suboptimal) greedy optimization of the user fractions can be approached arbitrarily closely.

## IV. Channel Estimation and Non-Perfect CSIT

So far, we have assumed that the transmitter (cluster controller) has perfect CSIT. In this section we consider the typical operation of a *Frequency Division Duplexing* (FDD) system, where where the BSs in each cluster acquire their CSIT by broadcasting a set of orthogonal downlink pilot signals (one signal per jointly coordinated antenna), in order to enable the users to measure their downlink channels and feed back the estimated channel state information in some form [37]. The noisy CSIT obtained by the training and feedback procedure is used at the cluster controller to compute the LZFB precoding matrix, to select the active users, and allocate power and rate. We seek the characterization of the non-trivial tradeoff between the advantage of having a large number of jointly processed transmit antennas (large $\gamma$ and/or large $B$) and the overhead required for estimating the channels. We assume that the channels are constant over time-frequency blocks of size $WT$ complex dimensions, where $W$ denotes to the system coherence bandwidth (in Hz) and $T$ denotes the system coherence time (in sec.). These blocks are identified with the scheduling slots of our downlink system. For slot, $\gamma_p BN$ dimensions are dedicated to downlink training, in order to allow all users in the cluster to estimate the composite channel (i.e., the corresponding column of $\mathbf{H}$ in (4)) formed by $\gamma BN$ coefficients. Since the channel vectors are Gaussian, linear MMSE estimation is optimal with respect to the MSE criterion. A simple dimensionality argument shows that the MSE can be made arbitrarily small as $\sigma_k^2 \to 0$ (vanishing noise plus ICI) if and only if $\gamma_p \geq \gamma$. The ratio $\gamma_p/\gamma \geq 1$ denotes the "pilot dimensionality overhead", relative to the minimum number of pilots for which the MMSE vanishes as $\sigma_k^2 \to 0$.

Focusing on the estimation of a generic column of $\mathbf{H}$ in (4) corresponding to some user $j$ in group $k$ of the reference cluster, the channel model of downlink channel estimation based on the common pilots is given by

$$\mathbf{y}_k^{(j)} = \mathbf{T}\underline{\mathbf{h}}_k^{(j)} + \mathbf{z}_k^{(j)} \tag{42}$$

where $\mathbf{T}$ is a $\gamma_p BN \times \gamma BN$ training matrix with equal-energy orthogonal columns, corresponding to the training sequences sent in parallel from the $\gamma BN$ cluster antennas (notice that the vertical dimension corresponds to channel uses, and the horizontal dimension corresponds to the antennas), the vector $\underline{\mathbf{h}}_k^{(j)}$ is the corresponding channel vector, obtained by stacking the channel vectors (including their path coefficients) corresponding to the different BSs in the cluster, and $\mathbf{z}_k^{(j)}$ is a vector of i.i.d. $\mathcal{CN}(0,1)$ noise plus interference samples. As before, we index the BSs forming the reference cluster as $m = 1, \dots, B$ and the user groups as $k = 1, \dots, A$. With this notation, from (4) we have

$$\underline{\mathbf{h}}_k^{(j)} = \begin{bmatrix} \beta_{1,k} \mathbf{h}_{1,k}^{(j)} \\ \vdots \\ \beta_{B,k} \mathbf{h}_{B,k}^{(j)} \end{bmatrix} \tag{43}$$

where $\mathbf{h}_{i,k}^{(j)}$ denotes the $j$-th column of the block $\mathbf{H}_{i,k}$, with i.i.d. $\mathcal{CN}(0,1)$ elements.

The equal-energy and orthogonality condition on the columns of $\mathbf{T}$ yield that the total transmit power (energy per channel use) in the training phase is given by

$$\frac{1}{\gamma_p B N} \text{tr} \left( \mathbf{T}^\mathsf{H} \mathbf{T} \right) = \frac{\gamma}{\gamma_p} p \tag{44}$$

where we let $\mathbf{T}^\mathsf{H} \mathbf{T} = p\mathbf{I}$, and $p$ denotes the energy of the training sequences. Letting the total training power equal to the total cluster transmit power, we obtain

$$p = \frac{\gamma_p}{\gamma} \sum_{m=1}^{B} P_m$$

Noticing that

$$\text{Cov}(\underline{\mathbf{h}}_k^{(j)}) = \text{diag}(\beta_{1,k}^2 \mathbf{I}, \dots, \beta_{B,k}^2 \mathbf{I}) \stackrel{\triangle}{=} \mathbf{D}_k$$

has block-diagonal structure with diagonal blocks given by scaled $\gamma N \times \gamma N$ identity matrices, we immediately obtain the MMSE estimator of $\underline{\mathbf{h}}_k^{(j)}$ in the form

$$\widehat{\underline{\mathbf{h}}}_k^{(j)} = \mathbf{D}_k \left( \mathbf{I} + p\mathbf{D}_k \right)^{-1} \mathbf{T}^\mathsf{H} \mathbf{y}_k^{(j)} \tag{45}$$

with estimation error covariance given by

$$\boldsymbol{\Sigma}_k = \mathbf{D}_k - p\mathbf{D}_k \left( \mathbf{I} + p\mathbf{D}_k \right)^{-1} \mathbf{D}_k = \mathbf{D}_k \left( \mathbf{I} + p\mathbf{D}_k \right)^{-1} \tag{46}$$

The MMSE covariance matrix is also block diagonal, with scaled identities diagonal blocks, and it depends only on the user group index $k$ and not on the individual user in the group (this is expected, since the users in the same group are statistically equivalent).

From the well-known orthogonality condition of MMSE estimation and from joint Gaussianity, we have the canonical decomposition

$$\underline{\mathbf{h}}_k^{(j)} = \widehat{\underline{\mathbf{h}}}_k^{(j)} + \underline{\mathbf{e}}_k^{(j)} \tag{47}$$

where the estimator $\widehat{\underline{\mathbf{h}}}_k^{(j)}$ and the error $\underline{\mathbf{e}}_k^{(j)}$ are independent, and such that

$$\mathrm{Cov}(\widehat{\underline{\mathbf{h}}}_k^{(j)}) = \mathbf{D}_k - \mathbf{\Sigma}_k = p\mathbf{D}_k \left(\mathbf{I} + p\mathbf{D}_k\right)^{-1} \mathbf{D}_k \tag{48}$$

Eventually, we can write the channel matrix $\mathbf{H}$ in (4) in the form $\mathbf{H} = \widehat{\mathbf{H}} + \mathbf{E}$, where

$$\widehat{\mathbf{H}} = \begin{bmatrix} \widehat{\beta}_{1,1}\mathbf{H}_{1,1} & \cdots & \widehat{\beta}_{1,A}\mathbf{H}_{1,A} \\ \vdots & & \vdots \\ \widehat{\beta}_{B,1}\mathbf{H}_{B,1} & \cdots & \widehat{\beta}_{B,A}\mathbf{H}_{B,A} \end{bmatrix}, \tag{49}$$

with

$$\widehat{\beta}_{m,k} = \frac{\beta_{m,k}^2}{\sqrt{1/p + \beta_{m,k}^2}}, \tag{50}$$

and the blocks $\mathbf{H}_{m,k}$ are independent with i.i.d. $\mathcal{CN}(0,1)$ elements, and where $\mathbf{E}$ is independent of $\widehat{\mathbf{H}}$, and is given in the form

$$\mathbf{E} = \begin{bmatrix} \bar{\beta}_{1,1}\mathbf{E}_{1,1} & \cdots & \bar{\beta}_{1,A}\mathbf{E}_{1,A} \\ \vdots & & \vdots \\ \bar{\beta}_{B,1}\mathbf{E}_{B,1} & \cdots & \bar{\beta}_{B,A}\mathbf{E}_{B,A} \end{bmatrix}, \tag{51}$$

with

$$\bar{\beta}_{m,k} = \sqrt{\beta_{m,k}^2 - \widehat{\beta}_{m,k}^2} = \frac{\beta_{m,k}}{\sqrt{1 + p\beta_{m,k}^2}}, \tag{52}$$

and the blocks $\mathbf{E}_{m,k}$ are independent with i.i.d. $\mathcal{CN}(0,1)$ elements.

In a practical FDD system, the users should feed back their estimated channel on each time-frequency block, i.e., for each new observation. Several schemes have been proposed for closed-loop CSIT feedback, including codebook-based vector quantization, scalar quantization of the channel coefficients, and unquantized "analog" feedback. CSIT feedback takes place on the uplink, and can be performed by accessing the uplink channel in FDMA/TDMA, or exploiting the MIMO-MAC nature of the uplink in order to allow a number of users proportional to the number of receiving antennas to send their feedback signals simultaneously (see [37], [46], [47] and references therein). Analyzing the system in the presence of a specific feedback scheme is possible [37]. However, from the results in the above mentioned papers we know that a well-designed digital feedback scheme can achieve a quantization error that is negligible with respect to the downlink training estimation error. Furthermore, this can be done with a moderate use of the uplink feedback total capacity, provided that the number of users feeding back their CSIT is not too large (see for example the optimization tradeoff in [48]). For the sake of simplicity, here we assume an ideal genie-aided CSIT feedback that provides $\widehat{\mathbf{H}}$ directly to the centralized cluster controller at no additional costs, either in terms of rate or in terms of CSIT distortion. This provides a "best case" for any scheme based on explicit downlink training and CSIT feedback. In the next section, we propose a randomized scheduling scheme that pre-selects at each slot a subset of users, such that only these selected users need to feed back their CSIT, thus limiting the cost of CSIT feedback on the uplink capacity.

As said before, the cluster controller computes a mismatched LZFB precoding matrix from the estimated channel matrix $\widehat{\mathbf{H}}$ instead of $\mathbf{H}$. The following theorem yields an achievability lower bound on the large-system performance of the mismatched LZFB:

*Theorem 3:* Under the downlink training scheme described above and assuming genie-aided CSIT feedback, the achievable rate of users in group $k$ is lower bounded by

$$R_k \geq \log\left(1 + \frac{\widehat{\Lambda}_k(\boldsymbol{\mu})q_k}{1 + \sum_{m=1}^{B}\bar{\beta}_{m,k}^2 P_m}\right), \tag{53}$$

where

$$\widehat{\Lambda}_k(\boldsymbol{\mu}) = \gamma\sum_{m=1}^{B}\widehat{\beta}_{m,k}^2\eta_m(\boldsymbol{\mu}), \tag{54}$$

where $(\eta_1(\boldsymbol{\mu}), \ldots, \eta_B(\boldsymbol{\mu}))$ is the unique solution with components in $[0, 1]$ of the fixed point equation

$$\eta_m = 1 - \sum_{q=1}^{A}\mu_q\frac{\eta_m\widehat{\beta}_{m,q}^2}{\gamma\sum_{\ell=1}^{B}\eta_\ell\widehat{\beta}_{\ell,q}^2}, \quad m = 1, \ldots, B, \tag{55}$$

with respect to the variables $\boldsymbol{\eta} = \{\eta_m\}$.

*Proof:* See Appendix D. ∎

It is immediate to see that all the derivations and the optimization made before for the case of perfect CSIT, including the system symmetry conditions, can be applied straightforwardly to the case of non-ideal CSIT, provided that the per-user rates are replaced by the corresponding terms in (53). In particular, Theorem 2 is valid by replacing $\{\beta_{m,k}\}$ with $\{\widehat{\beta}_{m,k}\}$ given in (50).

Finally, the system spectral efficiency must be scaled by the factor $\left[1 - \frac{\gamma_p NB}{WT}\right]_+$, that takes into account the downlink training overhead, i.e., fraction of dimensions per block dedicated to training. In particular, letting $\tau = \frac{N}{WT}$ denote the ratio between the number of users per group, $N$, and the dimensions in a time-frequency slot, we can investigate the system spectral efficiency for fixed $\tau$, in the limit of $N \to \infty$. The ratio $\tau$ captures the "dimensional crowding" of the system. It is clear that a highly underspread system ($WT \gg 1$) can accommodate more users and more jointly coordinated antennas at the transmitter. Vice versa, if $WT$ is not much larger than $N$, then the number of jointly coordinated transmit antennas (captured by the product $\gamma B$) is intrinsically limited by the channel time-frequency coherence.

## V. Numerical Results and Probabilistic Scheduling

In this section, first we provide a comparison between the large-system limit analytical results and the Monte Carlo simulation of the corresponding finite-dimensional systems with *greedy user selection*. Then, driven by the behavior of the finite-dimensional system, we propose a simplified probabilistic scheduling algorithm that randomly pre-select users according to the probability obtained from the asymptotic analysis. While greedy user selection needs that a large number of users (much larger than the scheduled active users) feed back their CSIT, in the proposed scheme the CSIT feedback is restricted just to the users that are effectively served. While a precise quantification the feedback resource requirements is out of the scope of this paper, we argue that the proposed scheme yields significant saving in the uplink feedback capacity, while achieving good throughput and fairness performance, when the system dimension is large. Finally, we consider the impact of non-perfect CSIT on the system performance and investigate the tradeoff between increasing the number of jointly coordinated antennas and the dimensionality cost incurrent by downlink training for channel estimation.

We consider a linear cellular arrangement where $M$ BSs are equally spaced on the segment $[-M, M]$ km, in positions $2m - M - 1$ for $m = 1, \ldots, M$, and $K$ user groups are also equally spaced on the same segment, with $K/M$ user groups uniformly spaced in each cell. The distance $d_{m,k}$ between BS $m$ and user group $k$ is defined modulo $[-M, M]$, i.e., we assume a wrap-around topology in order to eliminate boundary effects. We use a distance-dependent pathloss model given by $\alpha_{m,k}^2 = G_0/(1 + (d_{m,k}/\delta)^\nu))$ where the parameters $G_0, \nu$, and $\delta$ follow the mobile WiMAX system evaluation specifications [18], such that the 3dB break point is $\delta = 36$m (i.e., 3.6% of 1 km cell radius), the pathloss exponent is $\nu = 3.504$, the reference pathloss at $d_{m,k} = \delta$ is $G_0 = -91.64$ dB, and the per-BS transmit power normalized by the noise power at user terminals is $P = 154$ dB.

*1) Comparison with finite-dimensional systems:* Fig. 3 shows the average user throughputs (bit/s/Hz) versus user locations for the first two cells near the origin (given the symmetry, this pattern repeats periodically), for the case of $M = 8$ cells, $K = 64$ user groups, cluster size $B = 1, 2$ and 8 and $\gamma = 4$. Notice that with 8 user groups per cell and $\gamma = 4$, we have twice as much users as antennas in each cell. The case $B = 8$ corresponds to the network-wide full cooperation. For the finite-dimensional Monte Carlo simulation, we applied the same stochastic optimization algorithm described in Section III-D, where now $t$ denotes the scheduling slot index, and for each $t$ a new set of i.i.d. channel vectors is generated. In this case, the instantaneous weighted sum-rate is obtained via the user selection algorithm of [31], assuming that the CSIT for all users in the systems is available at the cluster controllers. As far as the network utility function $g(\cdot)$ is concerned, we consider the Proportional Fairness (PF) criterion, corresponding to $g(\overline{\mathbf{R}}) = \sum_k \log \overline{R}_k$. This PF criterion is applied to all the numerical results in this section.

From Fig. 3(a), we notice that the advantage of full cooperation is significant, whereas the cluster of size $B = 2$ yields a significant improvement for the users in the center of the cluster, with respect to the basic cellular system with no cluster cooperation ($B = 1$). In Fig. 3(b), we compare the asymptotic results with the finite-dimensional simulation results in the case of $B = 2$. The finite dimensional system yields better per-user throughput than the large-system limit, thanks to the ability of the user selection to exploit the randomness in the instantaneous channel realizations (multiuser diversity). However, as the number of users at each location, $N$ grows, the multiuser diversity gain continues to decline. For example, the relative gain of the finite-dimensional rate to the asymptotic rate is about 55% for $N = 1$ but only 25% for $N = 8$. In the case of $B = 1$ and $B = 8$ which are not shown here, the same trends are observed even though the diversity gain is slightly larger ($B = 1$) or smaller ($B = 8$). It is well-known that for large systems (large $N$), this multiuser diversity effect disappears because of "channel hardening" [33], [34].

*2) Random user pre-selection scheme for reduced CSIT feedback:* User selection requires a large amount of CSIT feedback since it needs CSIT from many users in order to select a good subset at each scheduling slot, even though no more users than the number of antennas can be served at a time. For systems with finite but large size, it is not wise to have many more users than transmit antennas to feedback their CSIT, since the multiuser diversity effect becomes marginal whereas the feedback resource grows at least linearly with the number of users feeding back their CSIT at each slot. In this regime, a meaningful option consists of pre-selecting the users to be served in each slot, such that only these users feed back their CSIT. In this case, we have to design a user pre-selection scheme that approximately maximizes the desired network utility function. For example, a simple round-robin scheme may perform far from the desired fairness optimal point.

For this purpose, we consider a probabilistic scheduling scheme based on our asymptotic analysis that effectively provides such user pre-selection. In the proposed scheme, the users to which CSIT feedback is requested are randomly selected in each slot $t$ as follows: let $\{\mu_k\}$ be the user fractions per group of (approximately) co-located users, which is obtained from the asymptotic analysis. The cluster controller has a maximum of $\gamma B N$ independent data streams to transmit using LZFB (equal to the number of jointly coordinated transmit antennas). At each slot $t$, the scheduler generates $\gamma B N$ i.i.d. random variables $S_1(t), \ldots, S_{\gamma B N}(t)$, taking values on the integers $\{0, 1, \ldots, A\}$ with probability $\mathbb{P}(S_i(t) = k) = \frac{\mu_k}{\gamma B}$ for $k \neq 0$ and $\mathbb{P}(S_i(t) = 0) = 1 - \sum_{k=1}^{A} \frac{\mu_k}{\gamma B}$. Then, user group $k$ is served by stream $i$ at slot $t$ if $S_i(t) = k$.

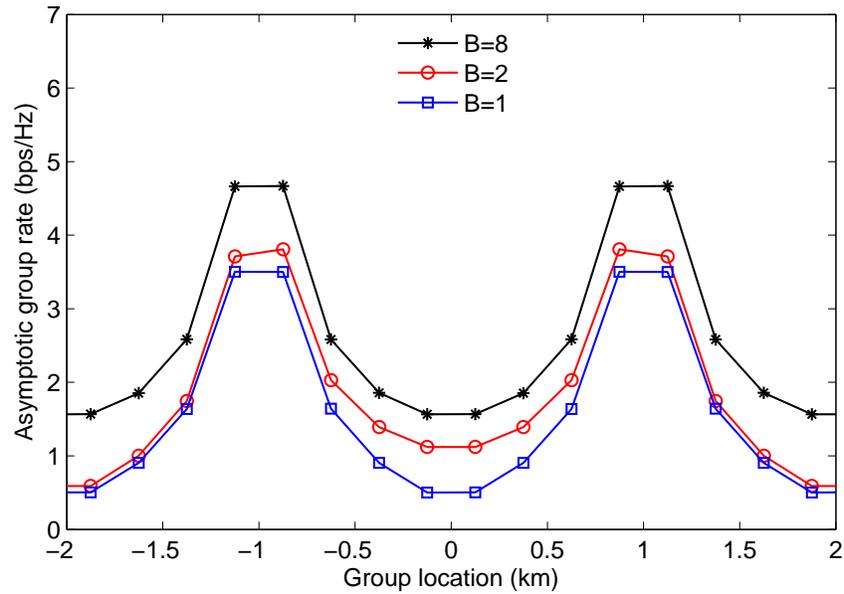

(a) Asymptotic analysis

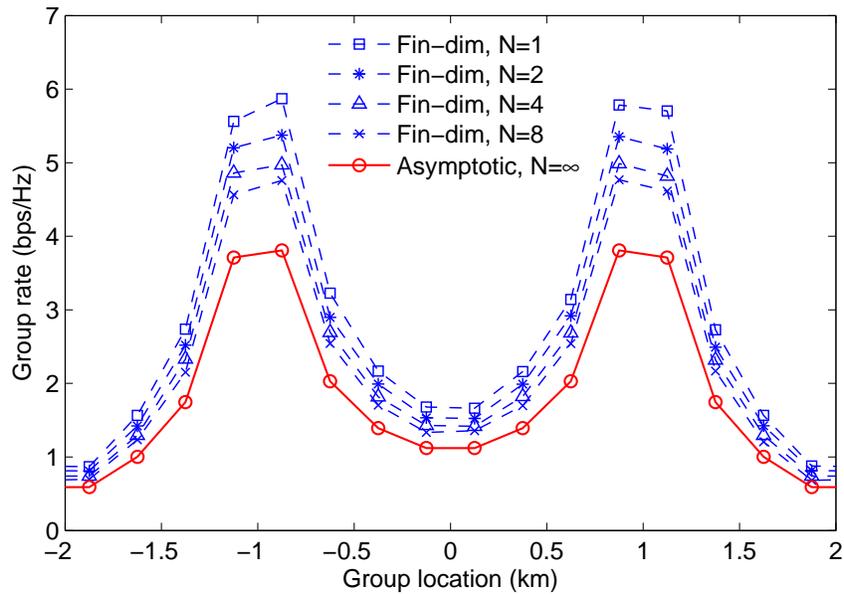

(b) Finite dimension simulation for $B = 2$

Fig. 3. User rate under perfect CSIT, obtained from asymptotic analysis for cooperation clusters of size $B = 1$ (no cooperation), 2, and 8 (full cooperation) and from finite dimension simulation with greedy user selection for $B = 2$ and $N = 1$, 2, 4, and 8. $M = 8$ cells and $K = 64$ user groups.

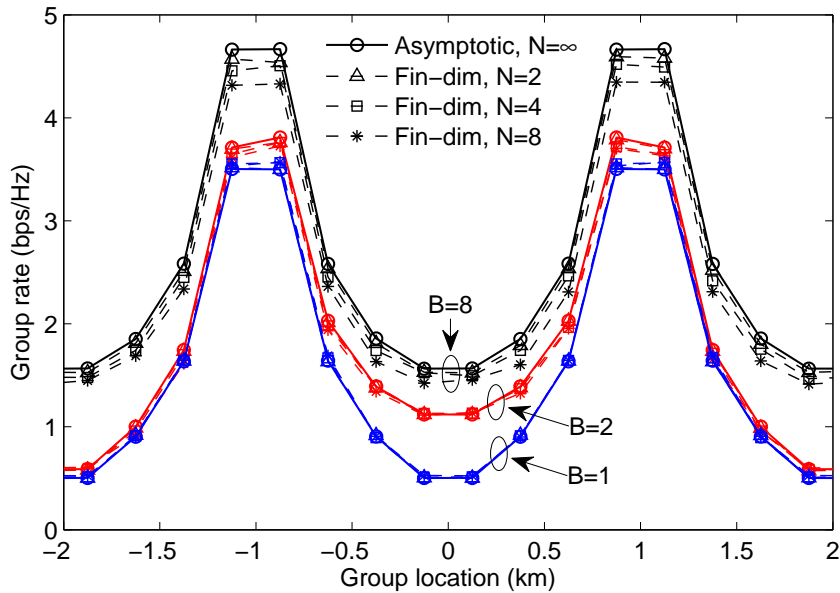

Fig. 4. User rate under perfect CSIT in finite dimension ($N = 2$, 4, and 8) with random user selection and power allocation aided by the asymptotic results for cooperation clusters of size $B$=1, 2, and 8. $M = 8$ cells and $K = 64$ user groups.

Notice that streams $i$'s for which $S_i(t) = 0$ are not used and that multiple streams may be associated to the same user group. Finally, for each stream a user in the associated group is selected at random, making sure that streams serve distinct users. Once the allocation of streams to users is determined, the selected users are requested to feedback their CSIT and the scheduler optimizes the transmit powers by solving the weighted sum rate maximization problem with weights $W_k = \partial g(\overline{\mathbf{R}})/\partial \overline{R}_k$, corresponding to the optimal asymptotic throughput point. In the special case of PF scheduling, this is given by $W_k = 1/\overline{R}_k$, [19].

The finite-dimension simulation results under this probabilistic user pre-selection scheme is compared with the asymptotic results in Fig. 4 under the same system setting as in Fig. 4. As $N$ increases, the finite-dimensional results converge to the infinite-dimensional limit and they are almost overlapped, especially when $B = 1$ or 2. Hence, the proposed scheme is effective for systems of finite but moderately large size.

*3) Non-perfect CSIT and coordination vs. estimation tradeoff:* Fig. 5 shows the cell sum rate (cluster sum rate normalized by the number of cooperating cells in the reference cluster) versus values of $\gamma$ in the cases of (a) perfect CSIT and no consideration of training overhead, and (b) non-perfect CSIT and explicit downlink training with $\gamma_p = \gamma$. We consider a larger number of user groups, $K = 192$ in the $M = 8$ cells. As shown in Fig. 5(a), under the assumption of perfect CSIT given at no cost, the cell sum rate grows almost linearly as $\gamma$ (the ratio of BS antennas over the users per group) increases, and grows also as $B$ (cluster size) increases, which shows the inter-cell cooperation and larger number of transmit antenna gain. However, when the CSIT estimation error and downlink training overhead are taken into account, there is a non-trivial tradeoff between the improvement owing to more and more jointly coordinated transmit antennas and the cost of estimating higher and higher dimensional channels.

Notice that this tradeoff is "fundamental", in the following sense: a trivial upper bound on the achievable sum capacity of the reference cluster is obtained by letting all users perfectly cooperate as a single multi-antenna receiver. The capacity of the resulting block-fading single-user MIMO channel with $\gamma BN$ transmit antennas and $AN$ receiving antennas and fading coherence block $WT = N/\tau$ was characterized in the high-SNR regime in [49], [50]. Using this result, in the case $\frac{1}{2\tau} \geq A \geq \gamma B$, the dimensionality "pre-log" loss factor with respect to the case of ideal CSIT is given by $\left(1 - \frac{\gamma BN}{WT}\right)$ that coincides with what is said at the end of Section IV with choice of $\gamma_p = \gamma$. In fact, the "pre-log" optimality of explicit training for single-user MIMO channels with block fading in the high-SNR regime is well-known [50], [51]. Also, the same result shows that if $\frac{1}{2\tau} < \min\{A, \gamma B\}$, then there is no point in using more than $WT/2$ jointly coordinated antennas. Finally, notice that the recently proposed schemes for "blind" interference alignment [52], exploiting reconfigurable antennas at the user terminals, still require channel state information at the receiver (CSIR) for coherent detection at each user terminal. Since the resulting channel is MIMO point-to-point, the same downlink training said above appears. In other words, these "blind" interference alignment schemes avoid CSIT feedback, but still require downlink training in the same amount considered in this work. In conclusions, while we have analyzed a specific downlink training scheme, we have that, for a cluster in isolation, the sum capacity scaling in the high-rate regime (high-SNR) is indeed the correct one.

Fig. 5 shows the cell sum rate with consideration of training overhead and estimation error for $\gamma_p = \gamma$. Inspired by practical system values, we chose $\tau = 1/64$ and $1/32$. In the finite-dimensional case, this corresponds to $WT = 640$ or $320$ signal dimensions, respectively, with $N = 10$ users per user groups (total $KN/M = 240$ users per cell). We notice that as $\gamma$ increases, the sum rates in most cases grow at first, achieve some maximum point and decrease, due to the tradeoff between the benefit from a large number of antennas and the training overhead cost. For given $B$ and $\tau$, the maximum sum rate is achieved at $\gamma B = \frac{1}{2\tau}$, which is in line with the result of the non-coherent MIMO high-SNR regime when $\gamma B \leq A$. For example, for $B = 2$ and $\tau = 1/64$, the sum rate is maximum at $\gamma = 16$ where $2\gamma = \frac{1}{2(1/64)}$. For $B = 1$ and $\tau = 1/64$, the optimal $\gamma$ is beyond the number of user groups per cell. We can also see

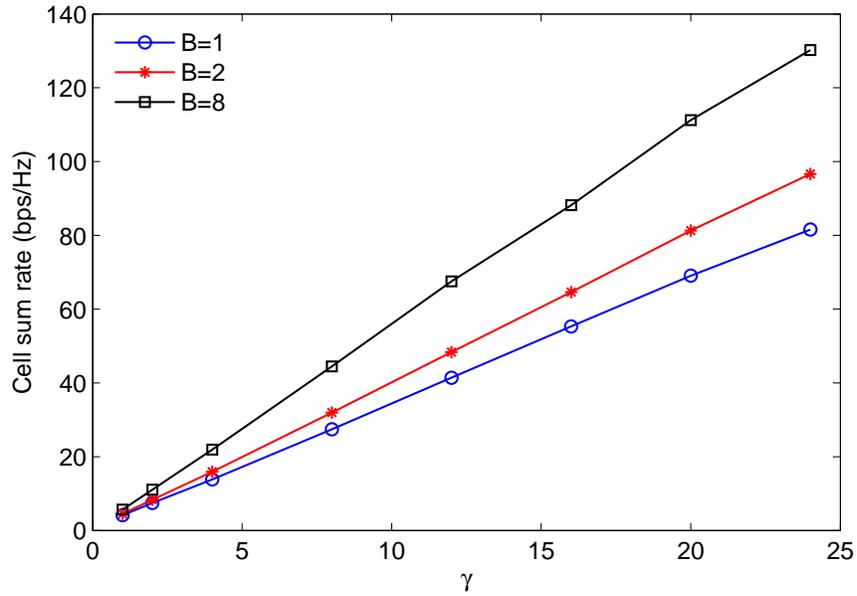

(a) Perfect CSIT and no training overhead

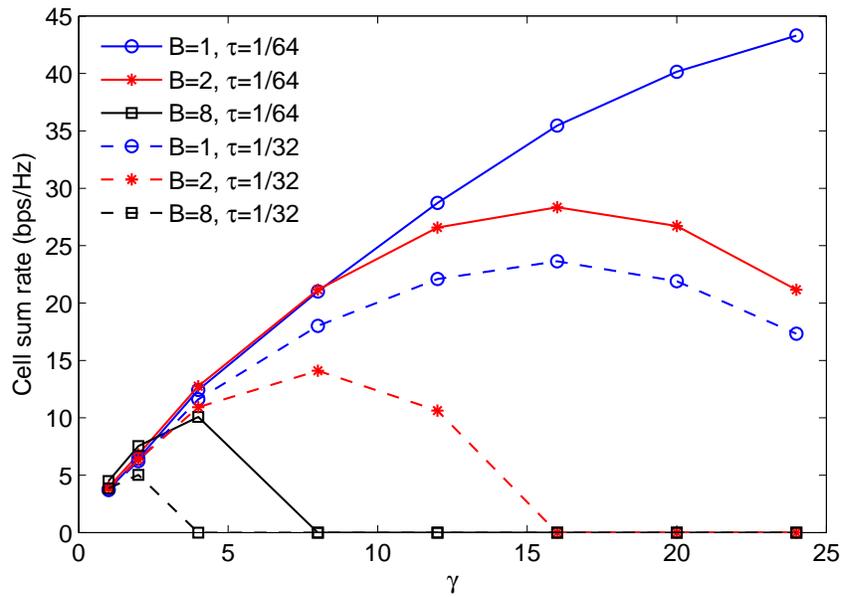

(b) Non-perfect CSIT and downlink training with $\gamma_p = \gamma$

Fig. 5. Cell sum rate versus antenna ratio $\gamma$ for cooperation clusters of size $B$=1, 2, and 8. $M = 8$ cells and $K = 192$ user groups.

that, when the number of antennas is large, the no cooperation case ($B = 1$) achieves the highest sum rate for both $\tau = 1/64$ and $1/32$, which suggests that no cooperation gain can be expected, because the improvement of multi-antenna gain does not compensate for the dimensional decrease (pre-log factor) due to the training overhead.

In order to see the best cluster size with downlink training and estimation, we consider a system with a large number of cells, $M = 24$. Fig. 6 illustrates the cell sum rate versus the cluster size $B$ for $\gamma = 1, 2, 4$, and $8$ and $\tau = 1/64$ and $1/32$ with $\gamma_p = \gamma$. In a linear cellular arrangement with $M = 24$, the clusters except for $B = 1, 2$, or $24$, do not have the symmetric structure described in Section III-A. So for those clusters, we notice that the solution of problem (36) under the cluster sum-power constraint produces an upper-bound of the optimal value under per-BS power constraint. Even though not explicitly shown in the figure, we can confirm that the cluster sum rate ($B$ times the cell sum rate) is maximized when $\gamma B = \frac{1}{2\tau}$ but as far as the cell sum rate is concerned, the optimal $\gamma$ for given $B$ and $\tau$ is smaller than the optimal one in terms of the cluster sum rate, i.e., $\frac{1}{2\tau\gamma}$. For example, in Fig. 6(a), when $\gamma = 4$ and $\tau = 1/64$, the maximum cluster sum rate is achieved at $B = 8$, but the cell sum rate given as the cluster sum rate divided by $B$ is maximum at $B = 3$. When the channel is more time or frequency selective ($\tau = 1/32$), the optimum cluster size is $B = 1$ (no BS cooperation), as shown in Fig. 6(b). Furthermore, the cell sum rate is more sensitive to the cluster size, when the number of antennas is larger.

## VI. CONCLUSIONS

We considered a multi-cell "network MIMO" system in a realistic cellular scenario, with inter-cell cooperation and fairness criteria. Specifically, we focused on linear zero-forcing beamforming combined with user selection. We derived the asymptotic expression in the large system limit and proposed an algorithm that computes the throughput point under an arbitrary fairness criterion, expressed by the maximization of a suitable concave and componentwise increasing network utility function over the region of achievable ergodic user rates. The proposed method handles the per-cluster sum-power constraint. We showed that under certain system symmetries, this coincides with the more stringent per-BS power constraint. In particular, the system symmetries make the analysis much simpler, as it allows for a closed-form solution of a fixed-point equation that characterized the zero-forcing beamforming performance. The fairness scheduling was applied in the form of stochastic network optimization. The proposed asymptotic analysis is computationally much more efficient than the Monte Carlo simulation. It also provides a good approximation of finite-dimensional systems, when the users are randomly selected according to the asymptotic user fraction in the large system limit. In particular, we proposed a probabilistic scheduling scheme that randomly associates users with downlink data streams according to probabilities obtained from the asymptotic analysis, and provides a good approximation of the optimal throughput point while requiring much less CSIT feedback resource.

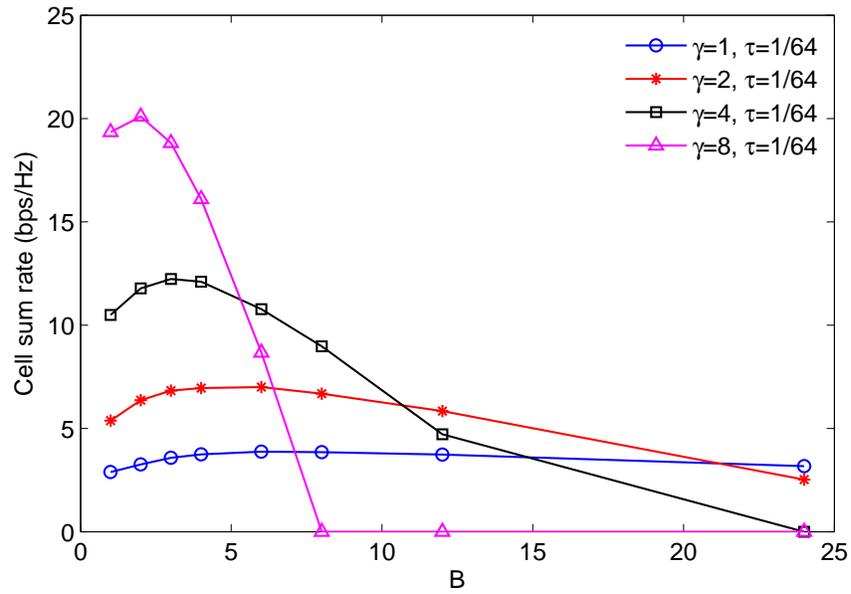

(a) $\tau = 1/64$

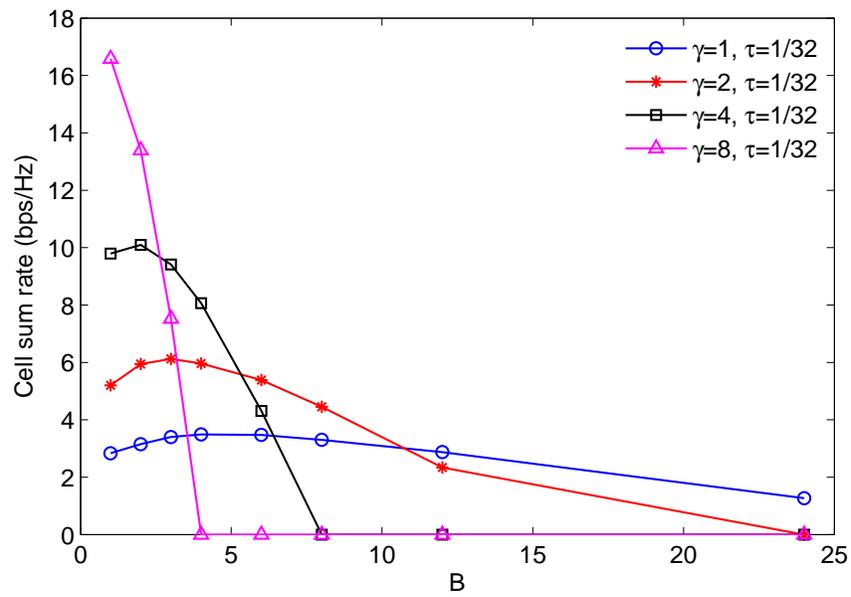

(b) $\tau = 1/32$

Fig. 6. Cell sum rate versus cluster size $B$ for antenna ratio $\gamma$=1, 2, 4, and 8 under non-perfect CSIT and explicit downlink training with $\gamma_p = \gamma$. $M = 24$ cells and $K = 192$ user groups.

Our analytic tool can be extended to handle explicit channel state information estimation, obtained from downlink training. This allows the investigation of the tradeoff between the number of jointly coordinated antennas and the cost of estimating higher dimensional channels. This tradeoff yields the optimal "cooperation cluster size" that maximizes the system throughput subject to fairness, when the cost of channel estimation is also taken into account. Due to this training overhead, the increase in the cooperation cluster size does not necessarily correspond to a system throughput increase. As a matter of fact, our analysis shows that in most cases no cooperation among base stations (conventional cellular systems) with a significant number of antennas per base station (large $\gamma$) yields the best system performance when the channel estimation cost is taken into account. This poses serious questions about whether "network MIMO" is a desirable solution, also taking into account that base station cooperation yields a non-trivial complexity increase in system implementation, requiring some form of centralized processing of all the $B$ base stations in each cluster. It clearly appear that more effort should be devoted to using a larger number of antennas at each base station, since this yields larger system throughput and significantly less system complexity.

## Appendix A

### Proof of Theorem 1

For the sake of clarity, we recall some definitions and facts about random matrices with independent non-identically distributed elements (see [53], [54]) of the type defined in (4), (7), that will be essential in the proof of Theorem 1.

*Definition 1:* Consider an $N_r \times N_c$ random matrix $\mathbf{H} = [\mathsf{H}_{i,j}]$, whose entries have variance

$$\text{Var}[\mathsf{H}_{i,j}] = \frac{\mathsf{P}_{i,j}}{N_r} \tag{56}$$

such that $\mathbf{P} = [\mathsf{P}_{i,j}]$ is an $N_r \times N_c$ deterministic matrix with uniformly bounded entries. For given $N_r$, we define the *variance profile* of $\mathbf{H}$ as the function $v^{N_r} : [0,1) \times [0,1) \to \mathbb{R}$ such that

$$v^{N_r}(x,y) = \mathsf{P}_{i,j}, \qquad \frac{i-1}{N_r} \leq x < \frac{i}{N_r}, \qquad \frac{j-1}{N_c} \leq y < \frac{j}{N_c} \tag{57}$$

■

When we consider the limit for $N_r \to \infty$ with fixed ratio $\frac{N_c}{N_r} \to \nu$, we assume that $v^{N_r}(x, y)$ converges uniformly to a bounded measurable function $v(x, y)$, referred to as the *asymptotic variance profile* of $\mathbf{H}$. For random matrices distributed according to Definition 1, we have the following results.

*Theorem 4 ( [53, Theorem 2.52]):* Let $\mathbf{H}$ be an $N_r \times N_c$ random matrix whose entries are independent zero-mean complex circularly symmetric random variables satisfying the Lindeberg condition

$$\frac{1}{N_r} \sum_{i,j} \mathbb{E}\left[|\mathsf{H}_{i,j}|^2 1\{|\mathsf{H}_{i,j}| \geq \delta\}\right] \to 0 \tag{58}$$

as $N_r \to \infty$ with $\frac{N_c}{N_r} \to \nu$, for all $\delta > 0$. Assume that the variances of the elements of $\mathbf{H}$ are given by Definition 1 and define the function

$$F^{(N_r)}(y, s) = \mathbf{h}_j^{\mathsf{H}} \left(\mathbf{I} + s \sum_{\ell \neq j} \mathbf{h}_\ell \mathbf{h}_\ell^{\mathsf{H}}\right)^{-1} \mathbf{h}_j, \qquad \frac{j-1}{N_c} \leq y < \frac{j}{N_c}. \tag{59}$$

As $N_r \to \infty$ with $\frac{N_c}{N_r} \to \nu$, $F^{(N_r)}(y, s)$ converges almost surely to the limit $F(y, s)$, given by the solution of the fixed-point equation

$$F(y, s) = \mathbb{E}\left[\frac{v(\mathsf{X}, y)}{1 + s\,\nu\,\mathbb{E}\left[\frac{v(\mathsf{X},\mathsf{Y})}{1+s\,F(\mathsf{Y},s)}\,\big|\,\mathsf{X}\right]}\right], \quad y \in [0, 1]. \tag{60}$$

where $\mathsf{X}$ and $\mathsf{Y}$ are i.i.d. random variables uniformly distributed on $[0, 1]$. ∎

Defining the effective dimension ratio as

$$\nu' = \nu \frac{\mathbb{P}\left(\mathbb{E}[v(\mathsf{X},\mathsf{Y})|\mathsf{Y}] \neq 0\right)}{\mathbb{P}\left(\mathbb{E}[v(\mathsf{X},\mathsf{Y})|\mathsf{X}] \neq 0\right)},$$

the following high-SNR limit can be proved.

*Corollary 1 (see [53, Theorem 3.1]):* As $s$ goes to infinity, we have

$$\lim_{s \to \infty} F(y, s) = \begin{cases} \Psi_\infty(y) & \text{if } \nu' < 1 \\ 0 & \text{if } \nu' \geq 1 \end{cases} \tag{61}$$

where, for $\nu' < 1$, $\Psi_\infty(y)$ is the positive solution to

$$\Psi_\infty(y) = \mathbb{E}\left[\frac{v(\mathsf{X}, y)}{1 + \nu \mathbb{E}\left[\frac{v(\mathsf{X},\mathsf{Y})}{\Psi_\infty(\mathsf{Y})}\,\big|\,\mathsf{X}\right]}\right] \tag{62}$$

∎

Now we enter specifically the proof of Theorem 1. Using the well-known formula for the inverse of a $2 \times 2$ block matrix, we can write the $(j, j)$ diagonal element of the matrix $(\mathbf{I} + s\mathbf{H}^{\mathsf{H}}\mathbf{H})^{-1}$ as

$$\left[\left(\mathbf{I} + s\mathbf{H}^{\mathsf{H}}\mathbf{H}\right)^{-1}\right]_{j,j} = \frac{1}{1 + s\mathbf{h}_j^{\mathsf{H}}\left(\mathbf{I} + s \sum_{\ell \neq j} \mathbf{h}_\ell \mathbf{h}_\ell^{\mathsf{H}}\right)^{-1} \mathbf{h}_j} \tag{63}$$

Furthermore, assuming that $\mathbf{H}$ has full rank, then

$$
\begin{aligned}
\left[ \left( \mathbf{H}^{\mathsf{H}} \mathbf{H} \right)^{-1} \right]_{j,j} &= \lim_{s \to \infty} s \left[ \left( \mathbf{I} + s \mathbf{H}^{\mathsf{H}} \mathbf{H} \right)^{-1} \right]_{j,j} \\
&= \lim_{s \to \infty} \frac{s}{1 + s \mathbf{h}_j^{\mathsf{H}} \left( \mathbf{I} + s \sum_{\ell \neq j} \mathbf{h}_\ell \mathbf{h}_\ell^{\mathsf{H}} \right)^{-1} \mathbf{h}_j} \\
&= \frac{1}{\displaystyle \lim_{s \to \infty} \mathbf{h}_j^{\mathsf{H}} \left( \mathbf{I} + s \sum_{\ell \neq j} \mathbf{h}_\ell \mathbf{h}_\ell^{\mathsf{H}} \right)^{-1} \mathbf{h}_j}
\end{aligned}
\tag{64}
$$

Comparing the definition of $\Lambda_k^{(i)}(\boldsymbol{\mu})$ in (11) with (64) and using Theorem 4 and Corollary 1, we have that the desired limiting value of $\Lambda_k^{(i)}(\boldsymbol{\mu})$ is given by $\Psi_\infty(y)$, evaluated at the corresponding value of $y$ such that $\frac{j-1}{N_c} \leq y < \frac{j}{N_c}$ for $j = \sum_{\ell=1}^{k-1} \mu_\ell N + i$, after replacing the general matrix $\mathbf{H}$ in Theorem 4 with $\mathbf{H}_{\boldsymbol{\mu}}$ given by our problem.

In this case, the number of rows in the matrix is given by $N_r = \gamma B N$ and the number of columns is given by $N_c = \mu N$. With the normalization by $1/\sqrt{N}$ of all the channel coefficients, the matrix $\mathbf{H}_{\boldsymbol{\mu}}$ defined in (7) is formed by independent blocks $\mathbf{H}_{m,k}(\mu_k)$ of dimension $\gamma N \times \mu_k N$, such that each block has i.i.d. $\mathcal{CN}(0, \beta_{m,k}^2/N)$ elements. As $N \to \infty$, we have that $N_c, N_r \to \infty$ with ratio $\nu = \frac{\mu}{\gamma B}$. By imposing the appropriate normalization, the asymptotic variance profile of $\mathbf{H}_{\boldsymbol{\mu}}$ is given by the piece-wise constant function

$$
v(x,y) = \gamma B \beta_{m,k}^2 \ \text{ for } \ (x,y) \in \left[ \frac{m-1}{B}, \frac{m}{B} \right) \times \left[ \frac{\mu_{1:k-1}}{\mu}, \frac{\mu_{1:k}}{\mu} \right)
\tag{65}
$$

with $m = 1, \dots, B$ and $k = 1, \dots, A$. Also, we find explicitly

$$
\nu' = \nu \, \frac{\sum_{k=1}^{A} \frac{\mu_k}{\mu} \mathbf{1} \left\{ \frac{1}{B} \sum_{m=1}^{B} \beta_{m,k} \neq 0 \right\}}{\frac{1}{B} \sum_{m=1}^{B} \mathbf{1} \left\{ \frac{1}{\mu} \sum_{k=1}^{A} \mu_k \beta_{m,k} \neq 0 \right\}}
\tag{66}
$$

and notice that the case $\nu' < 1$ in (61) always holds since, by construction, $\text{rank}(\mathbf{H}_{\boldsymbol{\mu}}) = \mu N$ almost surely. Hence, the limit for $\Lambda_k^{(i)}(\boldsymbol{\mu})$ is obtained as the solution of the fixed point equation (62), for any $y \in \left[ \frac{\mu_{1:k-1}}{\mu}, \frac{\mu_{1:k}}{\mu} \right)$. In fact, the piece-wise constant form of $v(x,y)$ yields that $\Lambda_k^{(i)}(\boldsymbol{\mu})$ converges to a limit that depends only on $k$ (the user group) and not on $i$ (the specific user in the group).

With some abuse of notation, we let $\Lambda_k(\boldsymbol{\mu}) = \Psi_\infty(y)$ for all $y \in \left[ \frac{\mu_{1:k-1}}{\mu}, \frac{\mu_{1:k}}{\mu} \right)$, in order to denote this limit. Particularizing (62) to this case, we obtain

$$
\Lambda_k(\boldsymbol{\mu}) = \gamma \sum_{m=1}^{B} \frac{\beta_{m,k}^2}{1 + \sum_{q=1}^{A} \mu_q \, \frac{\beta_{m,q}^2}{\Lambda_q(\boldsymbol{\mu})}}, \qquad k = 1, \dots, A
\tag{67}
$$

It follows that the asymptotic limit of $\mathbf{\Lambda_\mu}$ is block-diagonal, with scaled-identity diagonal blocks, where the $k$-th block is given by $\Lambda_k(\boldsymbol{\mu})\mathbf{I}_{\mu_k N}$.

The expression in (23) is eventually obtained by defining the auxiliary variables

$$\eta_m = \frac{1}{1 + \sum_{q=1}^{A} \mu_q \dfrac{\beta_{m,q}^2}{\Lambda_q(\boldsymbol{\mu})}}, \quad m = 1, \ldots, B, \tag{68}$$

such that, from (67), we have $\Lambda_k(\boldsymbol{\mu}) = \gamma \sum_{m=1}^{B} \beta_{m,k}^2 \eta_m$. Using this into (68), we obtain the fixed-point equations

$$\eta_m = \frac{1}{1 + \sum_{q=1}^{A} \dfrac{\mu_q \beta_{m,q}^2}{\gamma \sum_{\ell=1}^{B} \eta_\ell \beta_{\ell,q}^2}}, \quad m = 1, \ldots, B, \tag{69}$$

which, after simple manipulation, yields (24), as given in Theorem 1.

As a final remark, notice that (69) has some significant advantages with respect to (67). In particular, the variables $\eta_m$ take values in $[0, 1]$ (by construction), and typically we have $B < A$ (less BSs in a cluster than user groups). Therefore, (69) can be initialized by letting $\eta_m = 1$, and the fixed point equation iterative solution involves only $B$, rather than $A$, variables. Also, it is immediately evident by inspection that the solution of (69) for $\eta_m \in [0, 1]$ always exists and it is unique.

## Appendix B

## Proof of Theorem 2

Before entering the proof of Theorem 2, we state and prove two auxiliary results given here as Lemma 2 and Lemma 3.

*Lemma 2:* Let $\mathbf{H}$ be an $N_r \times N_c$ matrix of independent zero-mean elements whose variance is $O(1/N_r)$ and 4-th moment is $O(1/N_r^2)$, with converging variance profile as in Definition 1. Let $h_{i,j}$ denote its element in position $(i, j)$, $\mathbf{h}_j$ denote its $j$-th column, $\mathbf{r}_i$ denote its $i$-th row after removing the $(i, j)$-th element, and $\mathbf{c}_j$ denote its $j$-th column after removing the $(i, j)$-th element. Also, let $\mathbf{H}_j$ denote the matrix of dimension $N_r \times (N_c - 1)$ obtained from $\mathbf{H}$ by removing $\mathbf{h}_j$, and $\mathbf{H}_{i,j}$ denote the matrix of dimension $(N_r - 1) \times (N_c - 1)$ obtained from $\mathbf{H}_j$ by removing its $i$-th row, which coincides with $\mathbf{r}_i$. Consider the two quadratic forms:

$$\eta_j = \mathbf{h}_j^{\mathsf{H}} \left[ \mathbf{I} + s \mathbf{H}_j \mathbf{H}_j^{\mathsf{H}} \right]^{-1} \mathbf{h}_j \tag{70}$$

and

$$\eta_{i,j} = \mathbf{c}_j^{\mathsf{H}} \left[ \mathbf{I} + s \mathbf{H}_{i,j} \mathbf{H}_{i,j}^{\mathsf{H}} \right]^{-1} \mathbf{c}_j \tag{71}$$

where $s > 0$ is a real parameter and where $\mathbf{I}$ indicates an identity matrix of appropriate dimension. Assume that $\mathbf{H}$, $\mathbf{H}_j$ and $\mathbf{H}_{i,j}$ have all full column rank, i.e., $\mathrm{rank}(\mathbf{H}) = N_c$, $\mathrm{rank}(\mathbf{H}_j) = \mathrm{rank}(\mathbf{H}_{i,j}) = N_c - 1$, with probability 1. Then, for any $s > 0$, $N_c$ and $N_r \geq N_c$, the difference $\chi_{i,j} = \eta_j - \eta_{i,j}$ is a non-negative random variable that converges almost surely to zero as $N_r, N_c \to \infty$, when the ratio $N_c/N_r = \nu$ is kept constant.

*Proof:* Since the matrix $\mathbf{H}$ has independent elements with same order (with respect to $N_r$) of variance and 4-th moment, it is sufficient to prove the statement for $i = j = 1$. First, notice that $\eta_1$ and $\eta_{1,1}$ (defined through (70) and (71)) differ by the fact that the first row of the matrix $\mathbf{H}$ is removed. The lemma says, essentially, that under these conditions removing a row from the matrix $\mathbf{H}$ yields a very small variation of the value of the quadratic form (70), and indeed this variation tends to zero a.s. when the matrix dimensions grow large with fixed ratio. We can write

$$\mathbf{H} = \begin{bmatrix} h_{1,1} & \mathbf{r}_1 \\ \mathbf{c}_1 & \mathbf{H}_{1,1} \end{bmatrix}, \;\; \mathbf{H}_1 = \begin{bmatrix} \mathbf{r}_1 \\ \mathbf{H}_{1,1} \end{bmatrix}, \;\; \mathbf{h}_1 = \begin{bmatrix} h_{1,1} \\ \mathbf{c}_1 \end{bmatrix}. \tag{72}$$

In order to find a relationship between $\eta_1$ and $\eta_{1,1}$, we notice that

$$\left[ \mathbf{I} + s\mathbf{H}_1\mathbf{H}_1^{\mathsf{H}} \right] = \begin{bmatrix} 1 + s\mathbf{r}_1\mathbf{r}_1^{\mathsf{H}} & s\mathbf{r}_1\mathbf{H}_{1,1}^{\mathsf{H}} \\ s\mathbf{H}_{1,1}\mathbf{r}_1^{\mathsf{H}} & \mathbf{I} + s\mathbf{H}_{1,1}\mathbf{H}_{1,1}^{\mathsf{H}} \end{bmatrix} \tag{73}$$

Letting the 4 blocks of the $2 \times 2$ block matrix above be denoted by $\mathbf{M}_{11}, \mathbf{M}_{12}, \mathbf{M}_{21}, \mathbf{M}_{22}$ and defining the Schur complements of the diagonal blocks by

$$\mathbf{A}_{11} = \mathbf{M}_{11} - \mathbf{M}_{12}\mathbf{M}_{22}^{-1}\mathbf{M}_{21} = 1 + s\mathbf{r}_1\mathbf{r}_1^{\mathsf{H}} - s^2\mathbf{r}_1\mathbf{H}_{1,1}^{\mathsf{H}}\mathbf{M}_{22}^{-1}\mathbf{H}_{1,1}\mathbf{r}_1^{\mathsf{H}} \tag{74}$$

and

$$\mathbf{A}_{22} = \mathbf{M}_{22} - \mathbf{M}_{21}\mathbf{M}_{11}^{-1}\mathbf{M}_{12} = \mathbf{M}_{22} - \frac{s^2\mathbf{H}_{1,1}\mathbf{r}_1^{\mathsf{H}}\mathbf{r}_1\mathbf{H}_{1,1}^{\mathsf{H}}}{1 + s\mathbf{r}_1\mathbf{r}_1^{\mathsf{H}}} \tag{75}$$

we have the $2 \times 2$ matrix inversion formula

$$\left[ \mathbf{I} + s\mathbf{H}_1\mathbf{H}_1^{\mathsf{H}} \right]^{-1} = \begin{bmatrix} \mathbf{A}_{11}^{-1} & -\mathbf{A}_{11}^{-1}\mathbf{M}_{12}\mathbf{M}_{22}^{-1} \\ -\mathbf{A}_{22}^{-1}\mathbf{M}_{21}\mathbf{M}_{11}^{-1} & \mathbf{A}_{22}^{-1} \end{bmatrix}. \tag{76}$$

After some tedious but straightforward algebra based on repeated use of the Sherman-Morrison matrix inversion lemma (omitted for brevity), and noticing that in this case $\mathbf{A}_{11}$ is a scalar quantity, we arrive at

$$\begin{aligned} \eta_1 &= \mathbf{h}_1^{\mathsf{H}} \left[ \mathbf{I} + s\mathbf{H}_1\mathbf{H}_1^{\mathsf{H}} \right]^{-1} \mathbf{h}_1 \\ &= \frac{|h_{1,1}|^2 - 2s\mathrm{Re}\left\{ h_{1,1}^*\mathbf{r}_1\mathbf{H}_{1,1}^{\mathsf{H}}\mathbf{M}_{22}^{-1}\mathbf{c}_1 \right\} + s^2 \left| \mathbf{r}_1\mathbf{H}_{1,1}^{\mathsf{H}}\mathbf{M}_{22}^{-1}\mathbf{c}_1 \right|^2}{\mathbf{A}_{11}} + \mathbf{c}_1^{\mathsf{H}}\mathbf{M}_{22}^{-1}\mathbf{c}_1 \end{aligned} \tag{77}$$

From definition (71) we have that $\eta_{1,1} = \mathbf{c}_1^{\mathsf{H}} \mathbf{M}_{22}^{-1} \mathbf{c}_1$. Therefore,

$$
\begin{aligned}
\chi_{1,1} &= \eta_1 - \eta_{1,1} \\
&= \frac{|h_{1,1}|^2 - 2s\mathrm{Re}\left\{ h_{1,1}^* \mathbf{r}_1 \mathbf{H}_{1,1}^{\mathsf{H}} \mathbf{M}_{22}^{-1} \mathbf{c}_1 \right\} + s^2 \left| \mathbf{r}_1 \mathbf{H}_{1,1}^{\mathsf{H}} \mathbf{M}_{22}^{-1} \mathbf{c}_1 \right|^2}{\mathbf{A}_{11}} \qquad (78) \\
&= \frac{\left| h_{1,1} - s\mathbf{r}_1 \mathbf{H}_{1,1}^{\mathsf{H}} \mathbf{M}_{22}^{-1} \mathbf{c}_1 \right|^2}{\mathbf{A}_{11}} \qquad (79)
\end{aligned}
$$

For finite $N_r, N_c$, the RHS of the above equality is a non-negative random variable. First, we check that $\chi_{1,1}$ is well-defined for all $s > 0$. By writing explicitly the denominator, we have

$$
\begin{aligned}
\mathbf{A}_{11} &= 1 + s\mathbf{r}_1 \mathbf{r}_1^{\mathsf{H}} - s^2 \mathbf{r}_1 \mathbf{H}_{1,1}^{\mathsf{H}} \mathbf{M}_{22}^{-1} \mathbf{H}_{1,1} \mathbf{r}_1^{\mathsf{H}} \\
&= 1 + s\mathbf{r}_1 \mathbf{r}_1^{\mathsf{H}} - s^2 \mathbf{r}_1 \mathbf{H}_{1,1}^{\mathsf{H}} \left[ \mathbf{I} + s\mathbf{H}_{1,1} \mathbf{H}_{1,1}^{\mathsf{H}} \right]^{-1} \mathbf{H}_{1,1} \mathbf{r}_1^{\mathsf{H}} \\
&= 1 + s\mathbf{r}_1 \mathbf{r}_1^{\mathsf{H}} - s^2 \mathbf{r}_1 \mathbf{H}_{1,1}^{\mathsf{H}} \mathbf{H}_{1,1} \mathbf{r}_1^{\mathsf{H}} + s^2 \mathbf{r}_1 \mathbf{H}_{1,1}^{\mathsf{H}} \mathbf{H}_{1,1} \left[ \frac{1}{s} \mathbf{I} + \mathbf{H}_{1,1}^{\mathsf{H}} \mathbf{H}_{1,1} \right]^{-1} \mathbf{H}_{1,1}^{\mathsf{H}} \mathbf{H}_{1,1} \mathbf{r}_1^{\mathsf{H}} \qquad (80) \\
&= 1 + \mathbf{r}_1 \left[ s\mathbf{I} - s^2 \mathbf{H}_{1,1}^{\mathsf{H}} \mathbf{H}_{1,1} + s^2 \mathbf{H}_{1,1}^{\mathsf{H}} \mathbf{H}_{1,1} \left[ \frac{1}{s} \mathbf{I} + \mathbf{H}_{1,1}^{\mathsf{H}} \mathbf{H}_{1,1} \right]^{-1} \mathbf{H}_{1,1}^{\mathsf{H}} \mathbf{H}_{1,1} \right] \mathbf{r}_1^{\mathsf{H}} \qquad (81) \\
&= 1 + \mathbf{r}_1 \mathbf{U} \left[ s\mathbf{I} - s^2 \boldsymbol{\Gamma} + s^2 \boldsymbol{\Gamma} \left[ \frac{1}{s} \mathbf{I} + \boldsymbol{\Gamma} \right]^{-1} \boldsymbol{\Gamma} \right] \mathbf{U}^{\mathsf{H}} \mathbf{r}_1^{\mathsf{H}} \qquad (82) \\
&= 1 + \mathbf{r}_1 \mathbf{U} \, \mathrm{diag} \left( s - s^2 \Gamma_i + s^2 \Gamma_i^2 \frac{s}{1 + s\Gamma_i} \right) \mathbf{U}^{\mathsf{H}} \mathbf{r}_1^{\mathsf{H}} \\
&= 1 + \mathbf{r}_1 \mathbf{U} \, \mathrm{diag} \left( \frac{1}{1/s + \Gamma_i} \right) \mathbf{U}^{\mathsf{H}} \mathbf{r}_1^{\mathsf{H}} \qquad (83)
\end{aligned}
$$

where (80) follows by applying the matrix inversion lemma, (81) is obtained by collecting $\mathbf{r}_1$ on the left and on the right, (82) follows by letting $\mathbf{H}_{1,1}^{\mathsf{H}} \mathbf{H}_{1,1} = \mathbf{U} \boldsymbol{\Gamma} \mathbf{U}^{\mathsf{H}}$, where $\mathbf{U}$ is $(N_c - 1) \times (N_c - 1)$ unitary and $\boldsymbol{\Gamma}$ is the diagonal matrix of the eigenvalues of $\mathbf{H}_{1,1}^{\mathsf{H}} \mathbf{H}_{1,1}$, denoted by $\Gamma_i$, that are all strictly positive by assumption that $\mathrm{rank}(\mathbf{H}_{1,1}) = N_c - 1$, and finally (83) follows by simplifying the terms in the inner diagonal matrix. Notice that as $s \to \infty$, $\mathbf{A}_{11} \to 1 + \mathbf{r}_1 \left( \mathbf{H}_{1,1}^{\mathsf{H}} \mathbf{H}_{1,1} \right)^{-1} \mathbf{r}_1^{\mathsf{H}}$ where the convergence is monotone from below, i.e.,

$$
1 \leq \mathbf{A}_{1,1} \leq 1 + \mathbf{r}_1 \left( \mathbf{H}_{1,1}^{\mathsf{H}} \mathbf{H}_{1,1} \right)^{-1} \mathbf{r}_1^{\mathsf{H}}
$$

for all $s > 0$.

Next, we examine the term $s\mathbf{r}_1 \mathbf{H}_{1,1}^{\mathsf{H}} \mathbf{M}_{22}^{-1} \mathbf{c}_1$ in the numerator of (79). By proceeding in a very similar way as before, we arrive at (details are omitted for the sake of brevity):

$$
s\mathbf{r}_1 \mathbf{H}_{1,1}^{\mathsf{H}} \mathbf{M}_{22}^{-1} \mathbf{c}_1 = \mathbf{r}_1 \mathbf{U} \, \mathrm{diag} \left( \frac{1}{1/s + \Gamma_i} \right) \mathbf{U}^{\mathsf{H}} \mathbf{H}_{1,1}^{\mathsf{H}} \mathbf{c}_1 \qquad (84)
$$

For $s \to \infty$, we have that this term tends to $\mathbf{r}_1 \left( \mathbf{H}_{1,1}^{\mathsf{H}} \mathbf{H}_{1,1} \right)^{-1} \mathbf{H}_{1,1}^{\mathsf{H}} \mathbf{c}_1$. Since $\mathbf{r}_1$ and $\mathbf{c}_1$ are vectors of independent components, mutually independent, and independent of $\mathbf{H}_{1,1}$, and since they have elements with mean 0, variance $O(1/N_r)$ and 4-th order moment $O(1/N_r^2)$, also the inner product in the right-hand side of (84) has variance and 4-th order moment with the same behavior, for all $s > 0$. For $N_r \to \infty$, $N_c/N_r = \nu$, we have $\mathbf{A}_{11}$ tends to a finite deterministic constant $d(s) \geq 1$, that can be calculated by standard results in random matrix theory. In contrast, the numerator $\left| h_{1,1} - s\mathbf{r}_1 \mathbf{H}_{1,1}^{\mathsf{H}} \mathbf{M}_{22}^{-1} \mathbf{c}_1 \right|^2$ of $\chi_{1,1}$ converges to 0 a.s., since

$$\left| h_{1,1} - s\mathbf{r}_1 \mathbf{H}_{1,1}^{\mathsf{H}} \mathbf{M}_{22}^{-1} \mathbf{c}_1 \right|^2 \leq 4 \max \left\{ \left| h_{1,1} \right|^2, \left| s\mathbf{r}_1 \mathbf{H}_{1,1}^{\mathsf{H}} \mathbf{M}_{22}^{-1} \mathbf{c}_1 \right|^2 \right\}$$

and each of these terms tends to 0 a.s.. This concludes the proof of Lemma 2. ∎

*Remark 1:* We wish to remark at this point that the result of Lemma 2 is expected. In fact, notice that the two quadratic forms in (70) and (71) correspond to the "multiuser efficiency" of user $j$, in two systems that differ just by eliminating one row (the $i$-th row) from the spreading matrix $\mathbf{H}$. By eliminating a single row, the asymptotic properties of the matrix do not change. In fact, the variance profile and the matrix aspect ratio remain the same. Hence, it is completely expected that the difference between the two efficiencies, asymptotically, vanishes. A byproduct of the above analysis is that, since all quantities involved in the ratio (79) are bounded functions of $s$, by bounded convergence we can exchange the limits of $s \to \infty$ and $N_r \to \infty$. This will be used in the proof of Theorem B, since we notice that the limit of $\eta_1$ for $s \to \infty$ yields, for appropriate choice of the column $j$, the elements $\Lambda_k^{(i)}(\boldsymbol{\mu})$ (see (64)) appearing as the gains of the zero-forcing precoder. ◇

*Lemma 3:* Let $\mathbf{x}$ be a $n$-dimensional vector with i.i.d. entries with variance $\frac{1}{n}$. Let $\mathbf{A}$ and $\mathbf{C}$ be $n \times n$ Hermitian symmetric matrices independent of $\mathbf{x}$, and let $\mathbf{D}$ be a $n \times n$ diagonal matrix independent of $\mathbf{x}$. Then:

$$\mathbf{x}^{\mathsf{H}} \mathbf{D}^{\mathsf{H}} (\mathbf{D}\mathbf{x}\mathbf{x}^{\mathsf{H}}\mathbf{D}^{\mathsf{H}} + \mathbf{A})^{-1} \mathbf{C} (\mathbf{D}\mathbf{x}\mathbf{x}^{\mathsf{H}}\mathbf{D}^{\mathsf{H}} + \mathbf{A})^{-1} \mathbf{D}\mathbf{x} \to \frac{\phi(\mathbf{D}^{\mathsf{H}}\mathbf{A}^{-1}\mathbf{C}\mathbf{A}^{-1}\mathbf{D})}{(1 + \phi(\mathbf{D}^{\mathsf{H}}\mathbf{A}^{-1}\mathbf{D}))^2}$$

where $\phi(\cdot) = \lim_{n \to \infty} \frac{1}{n}\mathrm{tr}(\cdot)$ and the convergence is almost surely.

*Proof:* Let

$$Q = \mathbf{x}^{\mathsf{H}} \mathbf{D}^{\mathsf{H}} (\mathbf{D}\mathbf{x}\mathbf{x}^{\mathsf{H}}\mathbf{D}^{\mathsf{H}} + \mathbf{A})^{-1} \mathbf{C} (\mathbf{D}\mathbf{x}\mathbf{x}^{\mathsf{H}}\mathbf{D}^{\mathsf{H}} + \mathbf{A})^{-1} \mathbf{D}\mathbf{x}$$

From the inversion lemma we have that

$$\left( \mathbf{D}\mathbf{x}\mathbf{x}^{\mathsf{H}}\mathbf{D}^{\mathsf{H}} + \mathbf{A} \right)^{-1} = \mathbf{A}^{-1} - \frac{1}{1 + \mathbf{x}^{\mathsf{H}}\mathbf{D}^{\mathsf{H}}\mathbf{A}^{-1}\mathbf{D}\mathbf{x}} \mathbf{A}^{-1}\mathbf{D}\mathbf{x}\mathbf{x}^{\mathsf{H}}\mathbf{D}^{\mathsf{H}}\mathbf{A}^{-1} \tag{85}$$

Hence,

$$
\begin{aligned}
Q &= \mathbf{x}^{\mathsf{H}}\mathbf{D}^{\mathsf{H}}\mathbf{A}^{-1}\mathbf{C}(\mathbf{D}\mathbf{x}\mathbf{x}^{\mathsf{H}}\mathbf{D}^{\mathsf{H}} + \mathbf{A})^{-1}\mathbf{D}\mathbf{x} \\
&\quad - \frac{1}{1 + \mathbf{x}^{\mathsf{H}}\mathbf{D}^{\mathsf{H}}\mathbf{A}^{-1}\mathbf{D}\mathbf{x}}\mathbf{x}^{\mathsf{H}}\mathbf{D}^{\mathsf{H}}\mathbf{A}^{-1}\mathbf{D}\mathbf{x}\mathbf{x}^{\mathsf{H}}\mathbf{D}^{\mathsf{H}}\mathbf{A}^{-1}\mathbf{C}(\mathbf{D}\mathbf{x}\mathbf{x}^{\dagger}\mathbf{D}^{\dagger} + \mathbf{A})^{-1}\mathbf{D}\mathbf{x} \\
&= \frac{\mathsf{a}}{1 + \mathbf{x}^{\mathsf{H}}\mathbf{D}^{\mathsf{H}}\mathbf{A}^{-1}\mathbf{D}\mathbf{x}}
\end{aligned}
\tag{86}
$$

where

$$
\mathsf{a} = \mathbf{x}^{\mathsf{H}}\mathbf{D}^{\mathsf{H}}\mathbf{A}^{-1}\mathbf{C}(\mathbf{D}\mathbf{x}\mathbf{x}^{\mathsf{H}}\mathbf{D}^{\mathsf{H}} + \mathbf{A})^{-1}\mathbf{D}\mathbf{x}
$$

Applying again the inversion lemma we have:

$$
\begin{aligned}
\mathsf{a} &= \mathbf{x}^{\mathsf{H}}\mathbf{D}^{\mathsf{H}}\mathbf{A}^{-1}\mathbf{C}\mathbf{A}^{-1}\mathbf{D}\mathbf{x} \\
&\quad - \frac{1}{1 + \mathbf{x}^{\mathsf{H}}\mathbf{D}^{\mathsf{H}}\mathbf{A}^{-1}\mathbf{D}\mathbf{x}}\mathbf{x}^{\mathsf{H}}\mathbf{D}^{\mathsf{H}}\mathbf{A}^{-1}\mathbf{C}\mathbf{A}^{-1}\mathbf{D}\mathbf{x}\mathbf{x}^{\mathsf{H}}\mathbf{D}^{\mathsf{H}}\mathbf{A}^{-1}\mathbf{D}\mathbf{x} \\
&= \frac{\mathsf{b}}{1 + \mathbf{x}^{\mathsf{H}}\mathbf{D}^{\mathsf{H}}\mathbf{A}^{-1}\mathbf{D}\mathbf{x}}
\end{aligned}
\tag{87}
$$

where

$$
\mathsf{b} = \mathbf{x}^{\mathsf{H}}\mathbf{D}^{\mathsf{H}}\mathbf{A}^{-1}\mathbf{C}\mathbf{A}^{-1}\mathbf{D}\mathbf{x}
$$

From (86) and (87) we obtain

$$
Q = \frac{\mathbf{x}^{\mathsf{H}}\mathbf{D}^{\mathsf{H}}\mathbf{A}^{-1}\mathbf{C}\mathbf{A}^{-1}\mathbf{D}\mathbf{x}}{(1 + \mathbf{x}^{\mathsf{H}}\mathbf{D}^{\mathsf{H}}\mathbf{A}^{-1}\mathbf{D}\mathbf{x})^2}
\tag{88}
$$

Finally, we arrive at the desired result by using the well-known fact in random matrix theory according to which $\lim_{n\to\infty}\mathbf{x}^{\mathsf{H}}\mathbf{M}\mathbf{x} = \phi(\mathbf{M})$ provided that $\mathbf{x}, \mathbf{M}$ are independent, that $\mathbf{M}$ has a well-defined limiting eigenvalue distribution and that $\mathbf{x}$ has i.i.d. elements with mean zero and variance $1/n$. ∎

Using Lemmas 2 and 3, we can proceed with the proof Theorem 2. From the expression of $\theta_{m,k}(\boldsymbol{\mu})$, it follows that

$$
\begin{aligned}
\theta_{m,k}(\boldsymbol{\mu}) &= \frac{1}{N}\sum_{i=1}^{\mu_k N}\sum_{\ell=1+(m-1)\gamma N}^{m\gamma N}\left|\left[\mathbf{V}\boldsymbol{\mu}\right]_{\ell,k}^{(i)}\right|^2 \\
&= \frac{1}{N}\operatorname{tr}\left(\boldsymbol{\Phi}_m\mathbf{V}\boldsymbol{\mu}\boldsymbol{\Theta}_k\mathbf{V}_{\boldsymbol{\mu}}^{\mathsf{H}}\right) \\
&= \frac{1}{N}\operatorname{tr}\left(\boldsymbol{\Phi}_m\mathbf{H}\boldsymbol{\mu}(\mathbf{H}_{\boldsymbol{\mu}}^{\mathsf{H}}\mathbf{H}\boldsymbol{\mu})^{-1}\boldsymbol{\Lambda}_{\boldsymbol{\mu}}^{1/2}\boldsymbol{\Theta}_k\boldsymbol{\Lambda}_{\boldsymbol{\mu}}^{1/2}(\mathbf{H}_{\boldsymbol{\mu}}^{\mathsf{H}}\mathbf{H}\boldsymbol{\mu})^{-1}\mathbf{H}_{\boldsymbol{\mu}}^{\mathsf{H}}\boldsymbol{\Phi}_m\right)
\end{aligned}
\tag{89}
$$

where $\boldsymbol{\Phi}_m$ is a diagonal matrix with all zeros, but for $\gamma N$ consecutive ones, corresponding to positions from $(m-1)\gamma N+1$ to $m\gamma N$ on the main diagonal, and where $\boldsymbol{\Theta}_k$ denotes the $\mu N$-dimensional diagonal matrix with all zeros, but for $\mu_k N$ consecutive ones, corresponding to positions from $\mu_{1:k-1}N + 1$ to $\mu_{1:k}N$ on the main diagonal (recall that we define the partial sum $\mu_{1:k} = \sum_{j=1}^{k} \mu_j$).

The submatrix of $\mathbf{\Phi}_m \mathbf{H}_{\boldsymbol{\mu}}$ corresponding to the non-zero rows, i.e., including rows from $(m-1)\gamma N + 1$ to $m\gamma N$, can be written as

$$[\beta_{m,1}\mathbf{H}_{m,1}(\mu_1), \cdots, \beta_{m,A}\mathbf{H}_{m,A}(\mu_A)] = \mathbf{W}_m \mathbf{B}_m \tag{90}$$

where $\mathbf{W}_m$ is a $\gamma N \times \mu N$ rectangular matrix with i.i.d. entries, with mean 0 and variance $1/N$, and

$$\mathbf{B}_m = \text{diag}\left( \underbrace{\beta_{m,1}, \ldots, \beta_{m,1}}_{\mu_1 N}, \ldots, \underbrace{\beta_{m,k}, \ldots, \beta_{m,k}}_{\mu_k N}, \ldots, \underbrace{\beta_{m,A}, \ldots, \beta_{m,A}}_{\mu_A N} \right) \tag{91}$$

After simple algebraic manipulation, letting the $\ell$-th row of $\mathbf{W}_m$ be denoted by $\mathbf{w}_{m,\ell}^{\mathsf{H}}$ we can write

$$\begin{aligned}
\mathbf{H}_{\boldsymbol{\mu}}^{\mathsf{H}} \mathbf{H}_{\boldsymbol{\mu}} &= \sum_{m=1}^{B} \mathbf{B}_m \mathbf{W}_m^{\mathsf{H}} \mathbf{W}_m \mathbf{B}_m \\
&= \mathbf{B}_m \mathbf{w}_{m,\ell} \mathbf{w}_{m,\ell}^{\mathsf{H}} \mathbf{B}_m + \sum_{j \neq \ell} \mathbf{B}_m \mathbf{w}_{m,j} \mathbf{w}_{m,j}^{\mathsf{H}} \mathbf{B}_m \\
&\quad + \sum_{q \neq m} \mathbf{B}_q \mathbf{W}_q^{\mathsf{H}} \mathbf{W}_q \mathbf{B}_q.
\end{aligned} \tag{92}$$

In order to be able to apply Lemma 3, we need that the variance of the elements of the i.i.d. vector $\mathbf{w}_{m,\ell}$ (playing the role of $\mathbf{x}$ in the Lemma), is equal to the inverse of the vector length. Therefore, dividing by $\mu$, we define

$$\mathbf{A} = \frac{1}{\mu} \sum_{q=1}^{B} \mathbf{B}_q \mathbf{W}_q^{\mathsf{H}} \mathbf{W}_q \mathbf{B}_q \tag{93}$$

and

$$\mathbf{A}_{m,\ell} = \mathbf{A} - \frac{1}{\mu} \mathbf{B}_m \mathbf{w}_{m,\ell} \mathbf{w}_{m,\ell}^{\mathsf{H}} \mathbf{B}_m. \tag{94}$$

At this point, we let $\mathbf{\Lambda}_{\boldsymbol{\mu}}$ denote the diagonal matrix with diagonal element in position $(j, j)$ given by

$$\Lambda_k^{(i)}(\boldsymbol{\mu}) = \lim_{s \to \infty} \mathbf{h}_j^{\mathsf{H}} \left( \mathbf{I} + s \sum_{p \neq j} \mathbf{h}_p \mathbf{h}_p^{\mathsf{H}} \right)^{-1} \mathbf{h}_j$$

where $j = \mu_{1:k-1}N + i$, and where $\mathbf{h}_p$ indicates the $p$-th column of $\mathbf{H}_{\boldsymbol{\mu}}$. Also, we denote by $\widetilde{\mathbf{\Lambda}}_{\boldsymbol{\mu}}$ the diagonal matrix with diagonal element in position $(j, j)$ given by

$$\widetilde{\Lambda}_k^{(i)}(\boldsymbol{\mu}) = \lim_{s \to \infty} \widetilde{\mathbf{h}}_j^{\mathsf{H}} \left( \mathbf{I} + s \sum_{p \neq j} \widetilde{\mathbf{h}}_p \widetilde{\mathbf{h}}_p^{\mathsf{H}} \right)^{-1} \widetilde{\mathbf{h}}_j,$$

where $\widetilde{\mathbf{h}}_p$ denote the columns of $\mathbf{H}_{\boldsymbol{\mu}}$ after removing the row corresponding to $\mathbf{w}_{m,\ell}$, in the horizontal $m$-th "slice" of $\mathbf{H}_{\boldsymbol{\mu}}$ defined in (90). By the result of Lemma 2 and using the argument in Remark 1 about the exchange of the limits for $s \to \infty$ and for $N \to \infty$, we can write

$$\lim_{s \to \infty} \mathbf{h}_j^{\mathsf{H}} \left( \mathbf{I} + s \sum_{p \neq j} \mathbf{h}_p \mathbf{h}_p^{\mathsf{H}} \right)^{-1} \mathbf{h}_j = \lim_{s \to \infty} \widetilde{\mathbf{h}}_j^{\mathsf{H}} \left( \mathbf{I} + s \sum_{p \neq j} \widetilde{\mathbf{h}}_p \widetilde{\mathbf{h}}_p^{\mathsf{H}} \right)^{-1} \widetilde{\mathbf{h}}_j + \chi_j, \tag{95}$$

where $\chi_j$ is a non-negative random variable converging to zero almost surely, as $N \to \infty$.

We would like to apply Lemma 3 to a suitably manipulated version of the expression (89) of $\theta_{m,k}(\boldsymbol{\mu})$. As we shall see below, the vector $\mathbf{w}_{m,\ell}$ and a matrix that depends on the elements of $\boldsymbol{\Lambda}_{\boldsymbol{\mu}}$ play the role of $\mathbf{x}$ and $\mathbf{C}$ in Lemma 3. However, since $\mathbf{w}_{m,\ell}$ defines a row of $\mathbf{H}_{\boldsymbol{\mu}}$ and $\boldsymbol{\Lambda}_{\boldsymbol{\mu}}$ is also a function of $\mathbf{H}_{\boldsymbol{\mu}}$, the independent required by Lemma 3 does not hold. Therefore, we shall use $\widetilde{\boldsymbol{\Lambda}}_{\boldsymbol{\mu}}$ *in lieu* of $\boldsymbol{\Lambda}_{\boldsymbol{\mu}}$ in (89), allowing for a small error due to the terms $\chi_j$. By doing so, we will be able to apply Lemma 3 since $\mathbf{w}_{m,\ell}$ and $\widetilde{\boldsymbol{\Lambda}}_{\boldsymbol{\mu}}$ are statistically independent. By letting $N \to \infty$ and using Lemma 2, we shall show the error term vanishes almost surely.

Keeping in mind the above proof scheme, we write the matrix $\boldsymbol{\Lambda}_{\boldsymbol{\mu}}^{1/2} \boldsymbol{\Theta}_k \boldsymbol{\Lambda}_{\boldsymbol{\mu}}^{1/2}$ as:

$$\boldsymbol{\Lambda}_{\boldsymbol{\mu}}^{1/2} \boldsymbol{\Theta}_k \boldsymbol{\Lambda}_{\boldsymbol{\mu}}^{1/2} = \mathbf{C}_k^{(m,\ell)} + \boldsymbol{\Xi} \tag{96}$$

where both $\mathbf{C}_k^{(m,\ell)}$ are diagonal matrices, with

$$\left[ \mathbf{C}_k^{(m,\ell)} \right]_{j,j} = \begin{cases} \widetilde{\Lambda}_k^{(i)}(\boldsymbol{\mu}) & \text{for} \quad j = \mu_{1:k-1} N + i, \quad i = 1, \dots, N \\ 0 & \text{elsewhere} \end{cases} \tag{97}$$

and with

$$[\boldsymbol{\Xi}]_{j,j} = \begin{cases} \chi_j & \text{for} \quad j = \mu_{1:k-1} N + i, \quad i = 1, \dots, N \\ 0 & \text{elsewhere} \end{cases} \tag{98}$$

Notice that $\mathbf{B}_m$ and $\mathbf{C}_k^{(m,\ell)}$ have both dimension $\mu N \times \mu N$ and are independent of $\mathbf{w}_{m,\ell}$.

Using (91), (96), (93) and (94) in (89) we arrive at

$$
\begin{aligned}
\theta_{m,k}(\boldsymbol{\mu}) &= \frac{1}{N\mu}\mathrm{tr}\left(\frac{1}{\sqrt{\mu}}\mathbf{W}_m\mathbf{B}_m\mathbf{A}^{-1}\left(\mathbf{C}_k^{(m,\ell)}+\boldsymbol{\Xi}^{(m,\ell)}\right)\mathbf{A}^{-1}\mathbf{B}_m\mathbf{W}_m^{\mathsf{H}}\frac{1}{\sqrt{\mu}}\right)\\
&= \frac{1}{N\mu}\sum_{\ell=1}^{\gamma N}\frac{1}{\sqrt{\mu}}\mathbf{w}_{m,\ell}^{\mathsf{H}}\mathbf{B}_m\left(\frac{1}{\mu}\mathbf{B}_m\mathbf{w}_{m,\ell}\mathbf{w}_{m,\ell}^{\mathsf{H}}\mathbf{B}_m+\mathbf{A}_{m,\ell}\right)^{-1}\left(\mathbf{C}_k^{(m,\ell)}+\boldsymbol{\Xi}^{(m,\ell)}\right)\\
&\qquad\cdot\left(\frac{1}{\mu}\mathbf{B}_m\mathbf{w}_{m,\ell}\mathbf{w}_{m,\ell}^{\mathsf{H}}\mathbf{B}_m+\mathbf{A}_{m,\ell}\right)^{-1}\mathbf{B}_m\mathbf{w}_{m,\ell}\frac{1}{\sqrt{\mu}} \qquad\qquad (99)\\
&\rightarrow \frac{\gamma}{\mu}\frac{\phi\left(\mathbf{B}_m\mathbf{A}^{-1}\mathbf{C}_k\mathbf{A}^{-1}\mathbf{B}_m\right)}{\left(1+\phi\left(\mathbf{B}_m\mathbf{A}^{-1}\mathbf{B}_m\right)\right)^2} \qquad\qquad (100)
\end{aligned}
$$

$$
+\lim_{N\to\infty}\frac{1}{N\mu}\sum_{\ell=1}^{\gamma N}\frac{1}{\sqrt{\mu}}\mathbf{w}_{m,\ell}^{\mathsf{H}}\mathbf{B}_m\left(\frac{1}{\mu}\mathbf{B}_m\mathbf{w}_{m,\ell}\mathbf{w}_{m,\ell}^{\mathsf{H}}\mathbf{B}_m+\mathbf{A}_{m,\ell}\right)^{-1}\boldsymbol{\Xi}^{(m,\ell)}
$$
$$
\cdot\left(\frac{1}{\mu}\mathbf{B}_m\mathbf{w}_{m,\ell}\mathbf{w}_{m,\ell}^{\mathsf{H}}\mathbf{B}_m+\mathbf{A}_{m,\ell}\right)^{-1}\mathbf{B}_m\mathbf{w}_{m,\ell}\frac{1}{\sqrt{\mu}} \qquad\qquad (101)
$$

where the last line follows by applying Lemma 3, where we replaced $\mathbf{C}_k^{(m,\ell)}$ with $\mathbf{C}_k$ defined as in (97) and $\mathbf{A}_{m,\ell}$ with $\mathbf{A}$. This can be done since $\mathbf{C}_k^{(m,\ell)}$ and $\mathbf{C}_k$ have the same limit for $N\to\infty$ (see proof of Lemma 2), and by noticing that all the terms for different $\ell$ in the sum in (99) converge to the same limit (by statistical symmetry), that can be obtained by using $\mathbf{A}$ *in lieu* of $\mathbf{A}_{m,\ell}$ after applying Lemma 3.

Let's examine the limit of the error term in (101). We can write

$$
\frac{1}{N\mu}\sum_{\ell=1}^{\gamma N}\frac{1}{\sqrt{\mu}}\mathbf{w}_{m,\ell}^{\mathsf{H}}\mathbf{B}_m\left(\frac{1}{\mu}\mathbf{B}_m\mathbf{w}_{m,\ell}\mathbf{w}_{m,\ell}^{\mathsf{H}}\mathbf{B}_m+\mathbf{A}_{m,\ell}\right)^{-1}\boldsymbol{\Xi}^{(m,\ell)}
$$
$$
\cdot\left(\frac{1}{\mu}\mathbf{B}_m\mathbf{w}_{m,\ell}\mathbf{w}_{m,\ell}^{\mathsf{H}}\mathbf{B}_m+\mathbf{A}_{m,\ell}\right)^{-1}\mathbf{B}_m\mathbf{w}_{m,\ell}\frac{1}{\sqrt{\mu}} \qquad\qquad (102)
$$
$$
\leq \chi_{\mathsf{max}}^{(m,\ell)}\frac{1}{N\mu}\sum_{\ell=1}^{\gamma N}\frac{1}{\sqrt{\mu}}\|\mathbf{w}_{m,\ell}\|^2\frac{\max_k\beta_{m,k}^2}{\kappa} \qquad\qquad (103)
$$

where $\kappa$ is a constant bounded away from zero and proportional to the square of the non-zero minimum eigenvalue of $\mathbf{A}$ (notice that $\mathbf{A}$ is invertible by construction), and where $\chi_{\mathsf{max}}^{(m,\ell)}$ is the maximum of the elements $\chi_j^{(m,\ell)}$. For $N\to\infty$, we have that $\|\mathbf{w}_{m,\ell}\|^2$ converges almost surely to a finite deterministic limit (strong law of large numbers). Hence, $\lim_{N\to\infty}\frac{1}{N\mu}\sum_{\ell=1}^{\gamma N}\frac{1}{\sqrt{\mu}}\|\mathbf{w}_{m,\ell}\|^2\frac{\max_k\beta_{m,k}^2}{\kappa}$ converges almost surely to a finite constant, and using Lemma 2 we have that the limit in (101) is zero almost surely.

We have concluded that the sought limit for $N\to\infty$ of $\theta_{m,k}(\boldsymbol{\mu})$ is given by the expression in (100). Therefore, our goal is now to evaluate the two limit normalized traces in (100). We start by the term in the denominator , which is considerably simpler. We have

$$
\begin{aligned}
\phi\left(\mathbf{B}_m \mathbf{A}^{-1} \mathbf{B}_m\right) &= \lim_{N \to \infty} \frac{1}{\mu N} \operatorname{tr}\left(\mathbf{B}_m \mathbf{A}^{-1} \mathbf{B}_m\right) \\
&= \lim_{N \to \infty} \frac{1}{\mu N} \operatorname{tr}\left(\left(\frac{1}{\mu} \mathbf{H}_{\boldsymbol{\mu}}^{\mathsf{H}} \mathbf{H}_{\boldsymbol{\mu}}\right)^{-1} \mathbf{B}_m^2\right) \\
&= \lim_{N \to \infty} \frac{1}{N} \operatorname{tr}\left(\left(\mathbf{H}_{\boldsymbol{\mu}}^{\mathsf{H}} \mathbf{H}_{\boldsymbol{\mu}}\right)^{-1} \mathbf{B}_m^2\right) \\
&= \lim_{N \to \infty} \frac{1}{N} \sum_{k=1}^{A} \sum_{i=1}^{\mu_k N} \frac{\beta_{m,k}^2}{\Lambda_k^{(i)}(\boldsymbol{\mu})} \\
&= \sum_{k=1}^{A} \frac{\mu_k \beta_{m,k}^2}{\Lambda_k(\boldsymbol{\mu})}
\end{aligned}
\tag{104}
$$

where we used the fact that, by definition,

$$
\left[\left(\mathbf{H}_{\boldsymbol{\mu}}^{\mathsf{H}} \mathbf{H}_{\boldsymbol{\mu}}\right)^{-1}\right]_k^{(i)} = \frac{1}{\Lambda_k^{(i)}(\boldsymbol{\mu})}
$$

for the diagonal elements of $\left(\mathbf{H}_{\boldsymbol{\mu}}^{\mathsf{H}} \mathbf{H}_{\boldsymbol{\mu}}\right)^{-1}$ in position $\mu_{1:k-1} N + i$ for $i = 1, \ldots \mu_k N$, and the convergence result of Theorem 1. Comparing (104) with the expression of $\eta_m$ in (68) in Appendix A, we have that

$$
\sum_{k=1}^{A} \frac{\mu_k \beta_{m,k}^2}{\Lambda_k(\boldsymbol{\mu})} = \frac{1}{\eta_m(\boldsymbol{\mu})} - 1,
\tag{105}
$$

where $\{\eta_m(\boldsymbol{\mu}) : m = 1, \ldots, B\}$ are defined in Theorem 1 as the solutions of the fixed-point equation (24). Then, the denominator of (100) can be written as

$$
\left(1 + \phi\left(\mathbf{B}_m \mathbf{A}^{-1} \mathbf{B}_m\right)\right)^2 = \eta_m^{-2}(\boldsymbol{\mu})
\tag{106}
$$

Next, we consider the numerator of (100). For this purpose, let $\rho$ be a dummy non-negative real variable and consider the identity:

$$
\frac{-d}{d\rho} \operatorname{tr}\left(\left(\rho \mathbf{B}_m^2 + \mathbf{A}\right)^{-1} \mathbf{C}_k\right) = \operatorname{tr}\left(\mathbf{B}_m(\rho \mathbf{B}_m^2 + \mathbf{A})^{-1} \mathbf{C}_k(\rho \mathbf{B}_m^2 + \mathbf{A})^{-1} \mathbf{B}_m\right)
\tag{107}
$$

By almost-sure continuity of the trace in the left-hand side of (107) with respect to $\rho \geq 0$, it follows that the desired expression for the numerator of (100) can be calculated as

$$
\phi\left(\mathbf{B}_m \mathbf{A}^{-1} \mathbf{C}_k \mathbf{A}^{-1} \mathbf{B}_m\right) = \lim_{\rho \downarrow 0} \frac{-d}{d\rho} \phi\left(\left(\rho \mathbf{B}_m^2 + \mathbf{A}\right)^{-1} \mathbf{C}_k\right)
\tag{108}
$$

In order to compute the asymptotic normalized trace in (108), we use [53, Lemma 2.51], reported here for completeness.

*Lemma 4:* Let $\mathbf{H}$ be $N_r \times N_c$ of the type given in Definition 1, satisfying the same assumptions of Theorem 4. For any $a, b \in [0, 1]$ with $a < b$,

$$\frac{1}{N_r} \sum_{i=\lfloor aN_r \rfloor}^{\lfloor bN_r \rfloor} \left[ \left( s\mathbf{H}\mathbf{H}^{\mathsf{H}} + \mathbf{I} \right)^{-1} \right]_{i,i} \to \int_a^b \Gamma_{\mathbf{H}\mathbf{H}^{\mathsf{H}}}(x, s) \, dx \tag{109}$$

where $N_c/N_r \to \nu$ and where $\Gamma_{\mathbf{H}\mathbf{H}^{\mathsf{H}}}(x, s)$ and $\Upsilon_{\mathbf{H}\mathbf{H}^{\mathsf{H}}}(y, s)$ are functions defined implicitly by the fixed-point equation

$$\begin{aligned}
\Gamma_{\mathbf{H}\mathbf{H}^{\mathsf{H}}}(x, s) &= \frac{1}{1 + \nu s \mathbb{E}\left[ v(x, \mathsf{Y}) \Upsilon_{\mathbf{H}\mathbf{H}^{\mathsf{H}}}(\mathsf{Y}, s) \right]} \\
\Upsilon_{\mathbf{H}\mathbf{H}^{\mathsf{H}}}(y, s) &= \frac{1}{1 + s \mathbb{E}\left[ v(\mathsf{X}, y) \Gamma_{\mathbf{H}\mathbf{H}^{\mathsf{H}}}(\mathsf{X}, s) \right]}
\end{aligned} \tag{110}$$

for $(x, y) \in [0, 1] \times [0, 1]$, where $\mathsf{X}$ and $\mathsf{Y}$ are i.i.d. uniform-$[0, 1]$ RVs and where the variance profile function $v(x, y)$ was introduced in Definition 1. ∎

In order to use Lemma 4 we write

$$\begin{aligned}
\operatorname{tr}\left( \left( \rho\mathbf{B}_m^2 + \mathbf{A} \right)^{-1} \mathbf{C}_k \right) &= \operatorname{tr}\left( \left( \rho\mathbf{I} + \mathbf{B}_m^{-1}\mathbf{A}\mathbf{B}_m^{-1} \right)^{-1} \mathbf{B}_m^{-1}\mathbf{C}_k\mathbf{B}_m^{-1} \right) \\
&= \frac{1}{\rho} \operatorname{tr}\left( \left( \mathbf{I} + \frac{1}{\rho}\mathbf{B}_m^{-1}\mathbf{A}\mathbf{B}_m^{-1} \right)^{-1} \mathbf{B}_m^{-1}\mathbf{C}_k\mathbf{B}_m^{-1} \right)
\end{aligned} \tag{111}$$

Noticing that, by definition, $\mathbf{A} = \frac{1}{\mu}\mathbf{H}_{\boldsymbol{\mu}}^{\mathsf{H}}\mathbf{H}_{\boldsymbol{\mu}}$, we can identify the matrix $\frac{1}{\sqrt{\mu}}\mathbf{B}_m^{-1}\mathbf{H}_{\boldsymbol{\mu}}^{\mathsf{H}}$ with the matrix $\mathbf{H}$ of Lemma 4. In this case, $N_r = \mu N$ and $N_c = \gamma B N$. Using $\{\mathbf{B}_m\}$ and $\{\mathbf{W}_m\}$ defined before, we can write the block-matrix form

$$\mathbf{H}_{\boldsymbol{\mu}}^{\mathsf{H}} = \left[ \mathbf{B}_1\mathbf{W}_1^{\mathsf{H}}, \mathbf{B}_2\mathbf{W}_2^{\mathsf{H}}, \ldots, \mathbf{B}_B\mathbf{W}_B^{\mathsf{H}} \right]$$

so that

$$\mathbf{B}_m^{-1}\mathbf{H}_{\boldsymbol{\mu}}^{\mathsf{H}} = \left[ \mathbf{B}_m^{-1}\mathbf{B}_1\mathbf{W}_1^{\mathsf{H}}, \mathbf{B}_m^{-1}\mathbf{B}_2\mathbf{W}_2^{\mathsf{H}}, \ldots, \mathbf{B}_m^{-1}\mathbf{B}_B\mathbf{W}_B^{\mathsf{H}} \right]$$

It follows that the variance profile function of $\frac{1}{\sqrt{\mu}}\mathbf{B}_m^{-1}\mathbf{H}_{\boldsymbol{\mu}}^{\mathsf{H}}$ is given by

$$v_m(x, y) = \frac{\beta_{\ell,k}^2}{\beta_{m,k}^2}, \quad \text{for } (x, y) \in \left[ \frac{\mu_{1:k-1}}{\mu}, \frac{\mu_{1:k}}{\mu} \right) \times \left[ \frac{\ell-1}{B}, \frac{\ell}{B} \right) \tag{112}$$

Using this in Lemma 4 and letting $1/\rho = s$, we find

$$\frac{1}{\mu N} \sum_{i=\mu_{1:k-1}N+1}^{\mu_{1:k}N} \left[ \left( \mathbf{I} + s\mathbf{B}_m^{-1}\mathbf{A}\mathbf{B}_m^{-1} \right)^{-1} \right]_{i,i} \to \int_{\mu_{1:k-1}/\mu}^{\mu_{1:k}/\mu} \Gamma_m(x, s) \, dx \tag{113}$$

where $\Gamma_m(x, s)$ and $\Upsilon_m(y, s)$ are defined by

$$\begin{aligned}
\Gamma_m(x, s) &= \frac{1}{1 + \frac{\gamma Bs}{\mu}\mathbb{E}\left[ v_m(x, \mathsf{Y})\Upsilon_m(\mathsf{Y}, s) \right]} \\
\Upsilon_m(y, s) &= \frac{1}{1 + s\mathbb{E}\left[ v_m(\mathsf{X}, y)\Gamma_m(\mathsf{X}, s) \right]}
\end{aligned} \tag{114}$$

Noticing that $v_m(x, y)$ is piecewise constant (see (112)), we have that also the functions $\Gamma_m(x, s)$ and $\Upsilon_m(y, s)$ are piecewise constant. With some abuse of notation, we denote the values of these functions as $\{\Gamma_{m,q}(s), q = 1, \ldots, A\}$ and $\{\Upsilon_{m,\ell}(s), \ell = 1, \ldots, B\}$, respectively, we find that (114) can be re-written directly in terms of these values as

$$
\begin{aligned}
\Gamma_{m,q}(s) &= \frac{1}{1 + \frac{s}{\mu} \sum_{\ell=1}^{B} \frac{\gamma \beta_{\ell,q}^2}{\beta_{m,q}^2} \Upsilon_{m,\ell}(s)}, \quad \text{for } q = 1, \ldots, A \\
\Upsilon_{m,\ell}(s) &= \frac{1}{1 + \frac{s}{\mu} \sum_{q=1}^{A} \frac{\mu_q \beta_{\ell,q}^2}{\beta_{m,q}^2} \Gamma_{m,q}(s)} \quad \text{for } \ell = 1, \ldots, B
\end{aligned}
\tag{115}
$$

Finally, using (113) and (111) and noticing that the non-zero diagonal elements of $\mathbf{B}_m^{-1} \mathbf{C}_k \mathbf{B}_m^{-1}$ converge to the constant $\Lambda_k(\boldsymbol{\mu}) \beta_{m,k}^{-2}$, we arrive at:

$$
\phi\left( \left( \rho \mathbf{B}_m^2 + \mathbf{A} \right)^{-1} \mathbf{C}_k \right) = \frac{\mu_k}{\rho \mu} \Gamma_{m,k}(1/\rho) \Lambda_k(\boldsymbol{\mu}) \beta_{m,k}^{-2}
\tag{116}
$$

It turns out that it is convenient to define the new variables

$$
S_{m,q}(\rho) = \frac{1}{\rho \beta_{m,q}^2} \Gamma_{m,q}(1/\rho), \quad \text{and} \quad G_{m,\ell}(\rho) = \Upsilon_{m,\ell}(1/\rho)
$$

Therefore, we can rewrite (115) and (116) as

$$
S_{m,q}(\rho) = \frac{1}{\rho \beta_{m,q}^2 + \frac{\gamma}{\mu} \sum_{\ell=1}^{B} \beta_{\ell,q}^2 G_{m,\ell}(\rho)}, \quad \text{for } q = 1, \ldots, A
\tag{117}
$$

$$
G_{m,\ell}(\rho) = \frac{1}{1 + \frac{1}{\mu} \sum_{q=1}^{A} \mu_q \beta_{\ell,q}^2 S_{m,q}(\rho)}, \quad \text{for } \ell = 1, \ldots, B
\tag{118}
$$

$$
\phi\left( \left( \rho \mathbf{B}_m^2 + \mathbf{A} \right)^{-1} \mathbf{C}_k \right) = \frac{\mu_k}{\mu} \Lambda_k(\boldsymbol{\mu}) S_{m,k}(\rho)
\tag{119}
$$

Taking the derivative in (119), we have that the numerator of (100) can be obtained as:

$$
\begin{aligned}
\lim_{\rho \downarrow 0} \frac{-d}{d\rho} \phi\left( \left( \rho \mathbf{B}_m^2 + \mathbf{A} \right)^{-1} \mathbf{C}_k \right) &= \frac{\mu_k}{\mu} \Lambda_k(\boldsymbol{\mu}) \lim_{\rho \downarrow 0} \frac{-d}{d\rho} S_{m,k}(\rho) \\
&= \frac{\mu_k}{\mu} \Lambda_k(\boldsymbol{\mu}) \dot{S}_{m,k}(0)
\end{aligned}
\tag{120}
$$

where we define $\dot{S}_{m,k}(0) = \frac{-d}{d\rho} S_{m,k}(\rho)|_{\rho=0}$ and, for later use, $\dot{G}_{m,\ell}(0) = \frac{d}{d\rho} G_{m,\ell}(\rho)|_{\rho=0}$.

Next, we wish to find a fixed-point equation that yields directly $\dot{S}_{m,k}(0)$. By continuity, we can replace directly $\rho = 0$ into the fixed point equations after taking the derivatives. By doing so, from (117) and (118), we obtain:

$$
\dot{S}_{m,q}(0) = \frac{\beta_{m,q}^2 + \frac{\gamma}{\mu} \sum_{\ell=1}^{B} \beta_{\ell,q}^2 \dot{G}_{m,\ell}(0)}{\left( \frac{\gamma}{\mu} \sum_{\ell=1}^{B} \beta_{\ell,q}^2 G_{m,\ell}(0) \right)^2}, \quad \text{for } q = 1, \ldots, A
\tag{121}
$$

$$
\dot{G}_{m,\ell}(0) = \frac{\frac{1}{\mu} \sum_{q=1}^{A} \mu_q \beta_{\ell,q}^2 \dot{S}_{m,q}(0)}{\left( 1 + \frac{1}{\mu} \sum_{q=1}^{A} \mu_q \beta_{\ell,q}^2 S_{m,q}(0) \right)^2}, \quad \text{for } \ell = 1, \ldots, B
\tag{122}
$$

Also, the equations for $S_{m,q}(0)$ and $G_{m,\ell}(0)$, obtained by replacing $\rho = 0$ in (117), (118), read:

$$S_{m,q}(0) = \frac{1}{\frac{\gamma}{\mu} \sum_{\ell=1}^{B} \beta_{\ell,q}^2 G_{m,\ell}(0)}, \quad \text{for } q = 1, \ldots, A \tag{123}$$

$$G_{m,\ell}(0) = \frac{1}{1 + \frac{1}{\mu} \sum_{q=1}^{A} \mu_q \beta_{\ell,q}^2 S_{m,q}(0)}, \quad \text{for } \ell = 1, \ldots, B \tag{124}$$

Replacing (123) into (124), we obtain, for all $\ell = 1, \ldots, B$,

$$G_{m,\ell}(0) = \frac{1}{1 + \sum_{q'=1}^{A} \frac{\mu_{q'} \beta_{\ell,q'}^2}{\gamma \sum_{\ell'=1}^{B} \beta_{\ell',q'}^2 G_{m,\ell'}(0)}}. \tag{125}$$

By multiplying both sides by $\gamma \beta_{\ell,q}^2$ and summing over $\ell$, we find

$$U_{m,q} = \gamma \sum_{\ell=1}^{B} \frac{\beta_{\ell,q}^2}{1 + \sum_{q'=1}^{A} \frac{\mu_{q'} \beta_{\ell,q'}^2}{U_{m,q'}}}, \tag{126}$$

where we define $U_{m,q} = \gamma \sum_{\ell=1}^{B} \beta_{\ell,q}^2 G_{m,\ell}(0)$. Comparing the fixed point equation (126) with (67), we discover that $U_{m,q} = \Lambda_q(\boldsymbol{\mu})$, independent of $m$. Using this result in (123), we obtain

$$S_{m,q}(0) = \frac{\mu}{\Lambda_q(\boldsymbol{\mu})} \tag{127}$$

Using the definition of $U_{m,q}$, (121) can be written as,

$$\dot{S}_{m,q}(0) = \frac{\mu^2 \beta_{m,q}^2 + \mu \dot{U}_{m,q}}{\Lambda_q^2(\boldsymbol{\mu})}, \tag{128}$$

where, with some abuse of notation, we define $\dot{U}_{m,q} = \gamma \sum_{\ell=1}^{B} \beta_{\ell,q}^2 \dot{G}_{m,\ell}(0)$.

Multiplying both sides of (122) by $\gamma \beta_{\ell,q}^2$, using (128) and (127) and summing over $\ell$, we obtain

$$\dot{U}_{m,q} = \gamma \sum_{\ell=1}^{B} \beta_{\ell,q}^2 \frac{\frac{1}{\mu} \sum_{q'=1}^{A} \mu_{q'} \beta_{\ell,q'}^2 \dot{S}_{m,q'}(0)}{\left(1 + \frac{1}{\mu} \sum_{q'=1}^{A} \mu_{q'} \beta_{\ell,q'}^2 S_{m,q'}(0)\right)^2}$$

$$= \gamma \sum_{\ell=1}^{B} \beta_{\ell,q}^2 \frac{\frac{1}{\mu} \sum_{q'=1}^{A} \mu_{q'} \beta_{\ell,q'}^2 \frac{\mu^2 \beta_{m,q'}^2 + \mu \dot{U}_{m,q'}}{\Lambda_{q'}^2(\boldsymbol{\mu})}}{\left(1 + \sum_{q'=1}^{A} \frac{\mu_{q'} \beta_{\ell,q'}^2}{\Lambda_{q'}(\boldsymbol{\mu})}\right)^2} \tag{129}$$

$$= \gamma \mu \sum_{q'=1}^{A} \left[\sum_{\ell=1}^{B} \eta_\ell^2(\boldsymbol{\mu}) \beta_{\ell,q}^2 \beta_{\ell,q'}^2\right] \frac{\mu_{q'}}{\Lambda_{q'}^2(\boldsymbol{\mu})} \left(\beta_{m,q'}^2 + \frac{1}{\mu} \dot{U}_{m,q'}(\boldsymbol{\mu})\right) \tag{130}$$

where we have used again the identity (105) in the denominator of (129). Somehow surprisingly, we notice that (130) is a system of $A$ linear equations in the $A$ unknown $\{\dot{U}_{m,q} : q = 1, \ldots, A\}$. Therefore, this can be solved explicitly (although not in closed form in general). In particular, we define the $A \times A$ matrix

$$\mathbf{M} = \left[\sum_{\ell=1}^{B} \eta_\ell^2(\boldsymbol{\mu}) \mathbf{b}_\ell \mathbf{b}_\ell^{\mathsf{T}}\right] \text{diag}\left(\frac{\mu_1}{\Lambda_1^2(\boldsymbol{\mu})}, \ldots, \frac{\mu_A}{\Lambda_A^2(\boldsymbol{\mu})}\right) \tag{131}$$

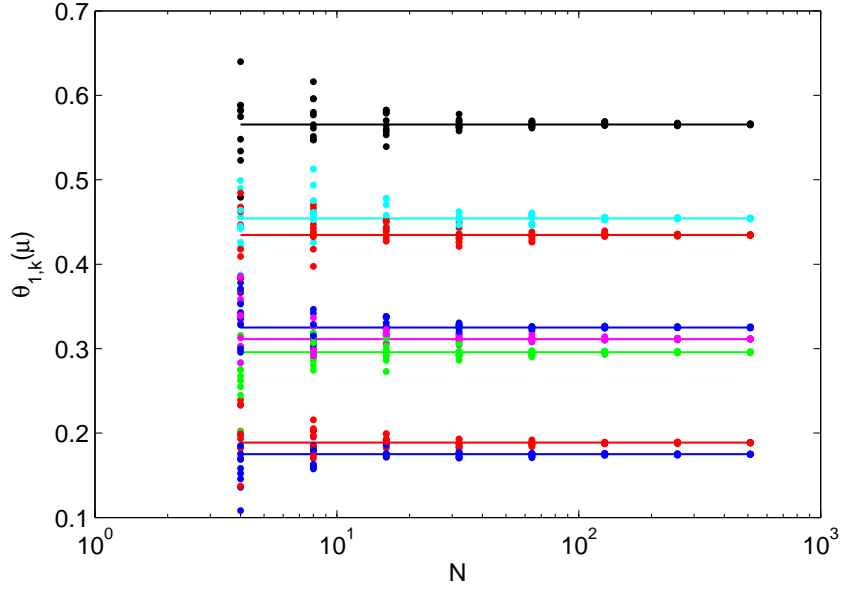

Fig. 7. Finite dimensional samples of $\theta_{m,k}(\boldsymbol{\mu})$ with $N = [4\ 8\ 16\ 32\ 64\ 128\ 256]$ (dots) and asymptotic values in the large system limit (lines) for $m = 1$, $k = 1, \ldots, 8$ under the settings of Example 1.

where $\mathbf{b}_\ell = (\beta_{\ell,1}^2, \ldots, \beta_{\ell,A}^2)^{\mathsf{T}}$, and the vector of unknowns $\dot{\mathbf{U}}_m$, then, we the linear system corresponding to (130) is given by

$$[\mathbf{I} - \gamma\mathbf{M}]\,\dot{\mathbf{U}}_m = \gamma\mu\mathbf{M}\mathbf{b}_m \tag{132}$$

Solving the system (132) and using (128) in (120), we obtain the sought numerator of (100) in the form

$$\frac{\mu_k}{\mu}\Lambda_k(\boldsymbol{\mu})\dot{S}_{m,k}(0) = \mu_k\frac{\mu\beta_{m,k}^2 + \dot{U}_{m,k}}{\Lambda_k(\boldsymbol{\mu})}. \tag{133}$$

Finally, putting together (100), (106) and (133), we obtain our final result:

$$
\begin{aligned}
\theta_{m,k}(\boldsymbol{\mu}) &= \frac{\gamma}{\mu}\frac{\phi\left(\mathbf{B}_m\mathbf{A}^{-1}\mathbf{C}_k\mathbf{A}^{-1}\mathbf{B}_m\right)}{\left(1 + \phi\left(\mathbf{B}_m\mathbf{A}^{-1}\mathbf{B}_m\right)\right)^2} \\
&= \frac{\gamma}{\mu}\frac{\mu_k(\mu\beta_{m,k}^2 + \dot{U}_{m,k})}{\Lambda_k(\boldsymbol{\mu})}\eta_m^2(\boldsymbol{\mu}) \\
&= \frac{\mu_k\eta_m^2(\boldsymbol{\mu})\left(\beta_{m,k}^2 + \dot{U}_{m,k}/\mu\right)}{\sum_{\ell=1}^{B}\eta_\ell(\boldsymbol{\mu})\beta_{\ell,k}^2}
\end{aligned}
\tag{134}
$$

where in the last line we used Theorem 1. Comparing (27) and (134), we see that the two expression coincide by letting $\xi_m = \dot{U}_m/\mu$. Therefore, Theorem 2 is proved.

Fig. 7 shows finite dimensional samples of $\theta_{m,k}(\boldsymbol{\mu})$ for randomly generated channels and their asymptotic values obtained from (134) under the settings of Example 1. As $N$ increases, the finite dimensional samples (dots) converge the asymptotic values (lines) and this example shows the validness of the asymptotic analysis.

We prove this lemma using the results of Theorem 2. Under the symmetric system conditions described in Section III-A1), each $\mathbf{b}_\ell = (\beta_{\ell,1}^2, \ldots, \beta_{\ell,A}^2)^\top$ is a cyclic shift of another by multiples of $A' = A/B$, i.e., $\beta_{m \oplus_B j,k} = \beta_{m,k \oplus_A jA'}$, as shown in (30). When $\mu_k = \mu_i'$ for all $k$ in equivalence class $i$, the matrix $\mathbf{M}$ in (29) becomes block-circulant with submatrices of size $A' \times A'$, since $\eta_\ell(\boldsymbol{\mu})$ is independent of $\ell$ and $\Lambda_k = \Lambda_i'$ for all $k$ in equivalence class $i$ under the symmetry conditions. Then, the matrix $[\mathbf{I} - \gamma \mathbf{M}]$, its inverse, and $[\mathbf{I} - \gamma \mathbf{M}]^{-1} \mathbf{M}$ are also block-circulant and multiplying a block-circulant matrix $[\mathbf{I} - \gamma \mathbf{M}]^{-1} \mathbf{M}$ by the cyclic shift of $\mathbf{b}_m$ by $jA'$ positions, for $j \in \mathbb{Z}$, produces the cyclic shift of $\boldsymbol{\zeta}_m$ in (28) by $jA'$ positions, i.e., we have $\zeta_{m \oplus_B j,k} = \zeta_{m,k \oplus_A jA'}$. From expression (27), noticing that $\mu_k = \mu_i'$ and $\sum_{\ell=1}^B \beta_{\ell,k}^2 = \beta_i^2$ are constants independent of $k$, for all $k$ in equivalence class $i$, and $\eta_m(\boldsymbol{\mu})$ is independent of $m$, we obtain that $\theta_{m \oplus_B j,k} = \theta_{m,k \oplus_A jA'}$, as we wanted to prove.

If also $q_k = q_i'$ for all $k$ in equivalence class $i$, we have

$$\sum_{k=1}^A q_k \theta_{m,k}(\boldsymbol{\mu}) = \sum_{\ell=1}^B \sum_{i=1}^{A'} q_i' \theta_{m,i \oplus_A \ell A'}(\boldsymbol{\mu}) = \sum_{i=1}^{A'} q_i' \sum_{\ell=1}^B \theta_{m \oplus_B \ell,i}(\boldsymbol{\mu}) = \sum_{i=1}^{A'} q_i' \mu_i'. \tag{135}$$

where we used the fact that, by construction of the matrix $\mathbf{V}_{\boldsymbol{\mu}}$ and the definition of the coefficients $\theta_{k,m}(\boldsymbol{\mu})$ (see (26)), the equality $\sum_{\ell=1}^B \theta_{k,\ell}(\boldsymbol{\mu}) = \mu_k$ holds in general (even in the non-symmetric case).

Let $\widehat{\mathbf{V}}_{\boldsymbol{\mu}}$ denote the beamforming matrix for given user fractions $\boldsymbol{\mu}$, defined as in Section II-B after replacing $\mathbf{H}_{\boldsymbol{\mu}}$ with $\widehat{\mathbf{H}}_{\boldsymbol{\mu}}$, defined as in (7) with the change $\beta_{m,k} \to \widehat{\beta}_{m,k}$.

Let's focus on a generic user $j$ in group $k$. From (3) and (9) the received signal is given by

$$\begin{aligned}
y_k^{(j)} &= \left(\underline{\mathbf{h}}_k^{(j)}\right)^{\mathsf{H}} \widehat{\mathbf{V}}_{\boldsymbol{\mu}} \mathbf{Q}^{1/2} \mathbf{u} + z_k^{(j)} \\
&= \left(\underline{\widehat{\mathbf{h}}}_k^{(j)}\right)^{\mathsf{H}} \widehat{\mathbf{v}}_k^{(j)} \sqrt{q_k^{(j)}} u_k^{(j)} + \left(\underline{\mathbf{e}}_k^{(j)}\right)^{\mathsf{H}} \widehat{\mathbf{V}}_{\boldsymbol{\mu}} \mathbf{Q}^{1/2} \mathbf{u} + z_k^{(j)}
\end{aligned} \tag{136}$$

where we used the fact that $\widehat{\mathbf{v}}_k^{(j)}$ is orthogonal to all measured channel vectors $\underline{\widehat{\mathbf{h}}}_\ell^{(i)}$, for all other scheduled users, and we used the decomposition (47). The useful signal coefficient $\left(\underline{\widehat{\mathbf{h}}}_k^{(j)}\right)^{\mathsf{H}} \widehat{\mathbf{v}}_k^{(j)}$ is, by construction, equal to the diagonal element corresponding to user $j$ in group $k$ of the matrix $\widehat{\boldsymbol{\Lambda}}_{\boldsymbol{\mu}}^{1/2}$, calculated from $\widehat{\mathbf{H}}\boldsymbol{\mu}$ as in (11). The additional interference term $\left(\underline{\mathbf{e}}_k^{(j)}\right)^{\mathsf{H}}\widehat{\mathbf{V}}\boldsymbol{\mu}\mathbf{u}$ is the intra-cluster multiuser interference due to the fact that CSIT is not perfect.

A standard technique to lower bound the mutual information $I(u_k^{(j)}; y_k^{(j)}|\widehat{\mathbf{H}})$ is as follows:

$$
\begin{aligned}
I(u_k^{(j)}; y_k^{(j)}|\widehat{\mathbf{H}}) &= h(u_k^{(j)}) - h(u_k^{(j)}|y_k^{(j)}, \widehat{\mathbf{H}}) \\
&= \log \pi e q_k - h(u_k^{(j)} - a y_k^{(j)}|y_k^{(j)}, \widehat{\mathbf{H}}) \\
&\geq \log \pi e q_k - h(u_k^{(j)} - a y_k^{(j)}|\widehat{\mathbf{H}}) \\
&\geq \log \pi e q_k - \mathbb{E}\left[\log \pi e \mathrm{Var}(u_k^{(j)} - a y_k^{(j)}|\widehat{\mathbf{H}})\right]
\end{aligned}
\tag{137}
$$

where we assumed that $u_k^{(j)}$ is Gaussian with variance 1 (denoting, as before, the transmit power to users in group $k$). The bound holds for any coefficient $a$. In particular, we wish to use the coefficient that minimizes the conditional variance $\mathrm{Var}(u_k^{(j)} - a y_k^{(j)}|\widehat{\mathbf{H}})$, given by the linear MMSE estimation of $u_k^{(j)}$ from $y_k^{(j)}$ for given $\widehat{\mathbf{H}}$. After standard algebra, omitted here for the sake of brevity, we obtain the variance (conditional MMSE estimation error)

$$
\mathrm{Var}(u_k^{(j)} - a y_k^{(j)}|\widehat{\mathbf{H}}) = \frac{q_k\left[\mathbb{E}\left[\left(\underline{\mathbf{e}}_k^{(j)}\right)^{\mathsf{H}}\widehat{\mathbf{V}}\boldsymbol{\mu}\mathbf{Q}\widehat{\mathbf{V}}_{\boldsymbol{\mu}}^{\mathsf{H}}\underline{\mathbf{e}}_k^{(j)}\Big|\widehat{\mathbf{H}}\right] + 1\right]}{\left|\left(\underline{\widehat{\mathbf{h}}}_k^{(j)}\right)^{\mathsf{H}}\mathbf{v}_k^{(j)}\right|^2 q_k + \mathbb{E}\left[\left(\underline{\mathbf{e}}_k^{(j)}\right)^{\mathsf{H}}\widehat{\mathbf{V}}\boldsymbol{\mu}\mathbf{Q}\widehat{\mathbf{V}}_{\boldsymbol{\mu}}^{\mathsf{H}}\underline{\mathbf{e}}_k^{(j)}\Big|\widehat{\mathbf{H}}\right] + 1}
\tag{138}
$$

Replacing this into (137), we obtain the desired lower bound in the form

$$
I(u_k^{(j)}; y_k^{(j)}|\widehat{\mathbf{H}}) \geq \mathbb{E}\left[\log\left(1 + \frac{\left|\left(\underline{\widehat{\mathbf{h}}}_k^{(j)}\right)^{\mathsf{H}}\mathbf{v}_k^{(j)}\right|^2 q_k}{1 + \mathbb{E}\left[\left(\underline{\mathbf{e}}_k^{(j)}\right)^{\mathsf{H}}\widehat{\mathbf{V}}\boldsymbol{\mu}\mathbf{Q}\widehat{\mathbf{V}}_{\boldsymbol{\mu}}^{\mathsf{H}}\underline{\mathbf{e}}_k^{(j)}\Big|\widehat{\mathbf{H}}\right]}\right)\right]
\tag{139}
$$

Let's examine the terms in (139) separately. As already said before, the coefficient in the numerator of the SINR term inside the logarithm, in the large system limit, is given by $\left|\left(\underline{\widehat{\mathbf{h}}}_k^{(j)}\right)^{\mathsf{H}}\mathbf{v}_k^{(j)}\right|^2 \to \widehat{\Lambda}_k(\boldsymbol{\mu})$, where the latter is obtained via Theorem 1 replacing the coefficients $\beta_{m,k}$ with the new coefficients $\widehat{\beta}_{m,k}$ defined in (50), thus obtaining (54) and (55).

The intra-cluster interference term in the denominator can be evaluated as follows. First, notice that because of the properties of the MMSE estimator, the channel error vector is independent of the estimator $\widehat{\mathbf{H}}$. Therefore, the conditioning with respect to $\widehat{\mathbf{H}}$ makes $\widehat{\mathbf{V}}\boldsymbol{\mu}$ and the diagonal matrix of transmitted powers $\mathbf{Q}$ act as constant matrices with respect to the conditional expectation, since they are both functions of the CSIT $\widehat{\mathbf{H}}$. We have

$$
\begin{aligned}
\mathbb{E}\left[\left.\left(\underline{\mathbf{e}}_k^{(j)}\right)^{\mathsf{H}}\widehat{\mathbf{V}}_{\boldsymbol{\mu}}\mathbf{Q}\widehat{\mathbf{V}}_{\boldsymbol{\mu}}^{\mathsf{H}}\underline{\mathbf{e}}_k^{(j)}\right|\widehat{\mathbf{H}}\right] &= \operatorname{tr}\left(\mathbf{Q}\widehat{\mathbf{V}}_{\boldsymbol{\mu}}^{\mathsf{H}}\operatorname{Cov}(\underline{\mathbf{e}}_k^{(j)})\widehat{\mathbf{V}}_{\boldsymbol{\mu}}\right) \\
&= \operatorname{tr}\left(\mathbf{Q}\widehat{\mathbf{V}}_{\boldsymbol{\mu}}^{\mathsf{H}}\boldsymbol{\Sigma}_k\widehat{\mathbf{V}}_{\boldsymbol{\mu}}\right) \\
&= \operatorname{tr}\left(\boldsymbol{\Sigma}_k\widehat{\mathbf{V}}_{\boldsymbol{\mu}}\mathbf{Q}\widehat{\mathbf{V}}_{\boldsymbol{\mu}}^{\mathsf{H}}\right) \\
&= \sum_{m=1}^{B}\bar{\beta}_{m,k}^2 P_m,
\end{aligned}
\tag{140}
$$

where the last line follows from the definition of $\boldsymbol{\Sigma}_k$ in (46), which is block-diagonal with $B$ diagonal blocks of dimension $\gamma N \times \gamma N$, and the $m$-th diagonal block is given by $\bar{\beta}_{m,k}^2\mathbf{I}$ where $\bar{\beta}_{m,k}$ is defined in (52), and by noticing that $\widehat{\mathbf{V}}_{\boldsymbol{\mu}}\mathbf{Q}\widehat{\mathbf{V}}_{\boldsymbol{\mu}}^{\mathsf{H}}$ is the covariance matrix of the signal transmitted from all the base stations forming the cluster. Under a per-BS power constraint, the partial trace of this matrix on any diagonal segment corresponding to base station $m$ (diagonal segments of length $\gamma N$) is equal to $P_m$. Therefore, the simple form of (140) follows. This concludes the proof.

## References

[1] IEEE 802.16 broadband wireless access working group, "IEEE 802.16m system requirements," IEEE 802.16m-07/002, Tech. Rep., Jan. 2010.

[2] 3GPP technical specification group radio access network, "Further advancements for E-UTRA: LTE-Advanced feasibility studies in RAN WG4," 3GPP TR 36.815, Tech. Rep., Mar. 2010.

[3] S. Parkvall, E. Dahlman, A. Furuskar, Y. Jading, M. Olsson, S. Wanstedt, and K. Zangi, "LTE-Advanced – Evolving LTE towards IMT-Advanced," in *Proc. IEEE Vehic. Tech. Conf. (VTC)*, Calgary, Alberta, Sept. 2008.

[4] J.-B. Landre, A. Saadani, and F. Ortolan, "Realistic performance of HSDPA MIMO in macro-cell environment," in *Proc. IEEE Int. Symp. on Personal, Indoor, and Mobile Radio Commun. (PIMRC)*, Tokyo, Japan, Sept. 2009.

[5] A. Farajidana, W. Chen, A. Damnjanovic, T. Yoo, D. Malladi, and C. Lott, "3GPP LTE downlink system performance," in *Proc. IEEE Global Commun. Conf. (GLOBECOM)*, Honolulu, HI, Nov. 2009.

[6] G. Caire and S. Shamai (Shitz), "On the achievable throughput of a multiantenna Gaussian broadcast channel," *IEEE Trans. on Inform. Theory*, vol. 49, no. 7, pp. 1691–1706, July 2003.

[7] P. Viswanath and D. N. C. Tse, "Sum capacity of the vector Gaussian broadcast channel and uplink-downlink duality," *IEEE Trans. on Inform. Theory*, vol. 49, no. 8, pp. 1912–1921, Aug. 2003.

[8] S. Vishwanath, N. Jindal, and A. Goldsmith, "Duality, achievable rates, and sum-rate capacity of Gaussian MIMO broadcast channels," *IEEE Trans. on Inform. Theory*, vol. 49, no. 10, pp. 2658–2668, Oct. 2003.

[9] W. Yu and J. M. Cioffi, "Sum capacity of Gaussian vector broadcast channels," *IEEE Trans. on Inform. Theory*, vol. 50, no. 9, pp. 1875–1892, Sept. 2004.

[10] H. Weingarten, Y. Steinberg, and S. Shamai (Shitz), "The capacity region of the Gaussian multiple-input multiple-output broadcast channel," *IEEE Trans. on Inform. Theory*, vol. 52, no. 9, pp. 3936–3964, Sept. 2006.

[11] G. J. Foschini, K. Karakayali, and R. A. Valenzuela, "Coordinating multiple antenna cellular networks to achieve enormous spectral efficiency," *IEE Proc. Commun.*, vol. 153, no. 4, pp. 548–555, Aug 2006.

[12] S. Jing, D. N. C. Tse, J. B. Soriaga, J. Hou, J. E. Smee, and R. Padovani, "Downlink macro-diversity in cellular networks," in *Proc. IEEE Int. Symp. on Inform. Theory (ISIT)*, Nice, France, June 2007.

[13] F. Boccardi and H. Huang, "Limited downlink network coordination in cellular networks," in *Proc. IEEE Int. Symp. on Personal, Indoor, and Mobile Radio Commun. (PIMRC)*, Athens, Greece, Sept. 2007.

[14] G. Caire, S. A. Ramprashad, H. C. Papadopoulos, C. Pepin, and C.-E. W. Sundberg, "Multiuser MIMO downlink with limited inter-cell cooperation: approximate interference alignment in time, frequency and space," in *Proc. Allerton Conf. on Commun., Control, and Computing*, Urbana-Champaign, IL, Sept. 2008.

[15] H. Dahrouj and W. Yu, "Coordinated beamforming for the multicell multi-antenna wireless system," *IEEE Trans. on Wireless Commun.*, vol. 9, no. 5, pp. 1748–1759, May 2010.

[16] H. Huh, H. C. Papadopoulos, and G. Caire, "Multiuser MISO transmitter optimization for intercell interference mitigation," *IEEE Trans. on Sig. Proc.*, vol. 58, no. 8, pp. 4272–4285, Aug. 2010.

[17] G. Boudreau, J. Panicker, N. Guo, R. Chang, N. Wang, and S. Vrzic, "Interference coordination and cancellation for 4G networks," *IEEE Commun. Mag.*, vol. 47, no. 4, pp. 74–81, Apr. 2009.

[18] WiMAX Forum, "Mobile WiMAX – Part I: A technical overview and performance evaluation," Tech. Rep., Aug. 2006.

[19] P. Viswanath, D. N. C. Tse, and R. Laroia, "Opportunistic beamforming using dumb antennas," *IEEE Trans. on Inform. Theory*, vol. 48, no. 6, pp. 1277–1294, June 2002.

[20] L. Georgiadis, M. J. Neely, and L. Tassiulas, *Resource Allocation and Cross-Layer Control in Wireless Networks*. Foundations and Trends in Networking, 2006, vol. 1, no. 1.

[21] H. Shirani-Mehr, G. Caire, and M. J. Neely, "MIMO downlink scheduling with non-perfect channel state knowledge," *IEEE Trans. on Commun.*, vol. 58, no. 7, pp. 2055–2066, July 2010.

[22] J. Mo and J. Walrand, "Fair end-to-end window-based congestion control," *IEEE/ACM Trans. on Networking*, vol. 8, pp. 556–567, Oct. 2000.

[23] H. Huang and R. A. Valenzuela, "Fundamental simulated performance of downlink fixed wireless cellular networks with multiple antennas," in *Proc. IEEE Int. Symp. on Personal, Indoor, and Mobile Radio Commun. (PIMRC)*, Berlin, Germany, Sept. 2005.

[24] J. Zhang, R. Chen, J. G. Andrews, A. Ghosh, and R. W. Heath, "Networked MIMO with clustered linear precoding," *IEEE Trans. on Wireless Commun.*, vol. 8, no. 4, pp. 1910–1921, Apr. 2009.

[25] H. Huang, M. Trivellato, A. Hottinen, M. Shafi, P. J. Smith, and R. A. Valenzuela, "Increasing downlink cellular throughput with limited network MIMO coordination," *IEEE Trans. on Wireless Commun.*, vol. 8, no. 6, pp. 2983–2989, June 2009.

[26] S. A. Ramprashad and G. Caire, "Cellular vs. network MIMO: a comparison including the channel state information overhead," in *Proc. IEEE Int. Symp. on Personal, Indoor, and Mobile Radio Commun. (PIMRC)*, Tokyo, Japan, Sept. 2009.

[27] H. Huh, S.-H. Moon, Y.-T. Kim, I. Lee, and G. Caire, "Multi-cell MIMO downlink with cell cooperation and fair scheduling: a large-system limit analysis," *submitted to IEEE Trans. on Inform. Theory*, June 2010. [Online]. Available: http://arxiv.org/abs/1006.2162

[28] G. Caire, S. A. Ramprashad, and H. C. Papadopoulos, "Rethinking network MIMO: cost of CSIT, performance Analysis, and architecture Comparisons," in *Proc. Inform. Theory and Appl. Workshop (ITA)*, San Diego, CA, Feb. 2010.

[29] A. M. Tulino and S. Verdu, "Asymptotic analysis of improved linear receivers for BPSK-CDMA subject to fading," *IEEE J. Select. Areas Commun.*, vol. 19, no. 8, pp. 1544–1555, Aug 2001.

[30] R. R. Muller and S. Verdu, "Design and analysis of low-complexity interference mitigation on vector channels," *IEEE J. Select. Areas Commun.*, vol. 19, no. 8, pp. 1429–1441, Aug 2001.

[31] G. Dimic and N. D. Sidiropoulos, "On downlink beamforming with greedy user selection: performance analysis and simple new algorithm," *IEEE Trans. on Sig. Proc.*, vol. 53, no. 10, pp. 3857–3868, Oct. 2005.

[32] T. Yoo and A. Goldsmith, "On the optimality of multiantenna broadcast scheduling using zero-forcing beamforming," *IEEE J. Select. Areas Commun.*, vol. 24, no. 3, pp. 528–541, Mar. 2006.

[33] B. M. Hochwald, T. L. Marzetta, and V. Tarokh, "Multiple-antenna channel hardening and its implications for rate feedback and scheduling," *IEEE Trans. on Inform. Theory*, vol. 50, no. 9, pp. 1893–1909, Sept. 2004.

[34] A. Tomasoni, G. Caire, M. Ferrari, and S. Bellini, "On the selection of semi-orthogonal users for zero-forcing beamforming," in *Proc. IEEE Int. Symp. on Inform. Theory (ISIT)*, Seoul, Korea, June 2009.

[35] P. Ding, D. J. Love, and M. D. Zoltowski, "Multiple antenna broadcast channels with shape feedback and limited feedback," *IEEE Trans. on Sig. Proc.*, vol. 55, no. 7, pp. 3417–3428, July 2007.

[36] J. Jose, A. Ashikhmin, P. Whiting, and S. Vishwanath, "Precoding methods for multi-user TDD MIMO systems," in *Proc. Allerton Conf. on Commun., Control, and Computing*, Urbana-Champaign, IL, Sept. 2008.

[37] G. Caire, N. Jindal, M. Kobayashi, and N. Ravindran, "Multiuser MIMO achievable rates with downlink training and channel state feedback," *IEEE Trans. on Inform. Theory*, vol. 56, no. 6, pp. 2845–2866, June 2010.

[38] S. Wagner, R. Couillet, M. Debbah, and D. T. M. Slock, "Large system analysis of linear precoding in MISO broadcast channels with limited feedback," *submitted to IEEE Trans. on Inform. Theory*, June 2009. [Online]. Available: http://arxiv.org/abs/0906.3682

[39] J. Hoydis, M. Kobayashi, and M. Debbah, "On the optimal number of cooperative base stations in network MIMO systems," *submitted to IEEE Trans. on Sig. Proc.*, Mar. 2010. [Online]. Available: http://arxiv.org/abs/1003.0332

[40] R. Zakhour and S. V. Hanly, "Base station cooperation on the downlink: large system analysis," *submitted to IEEE J. Select. Areas Commun.*, June 2010. [Online]. Available: http://arxiv.org/abs/1006.3360

[41] D. N. C. Tse and P. Viswanath, *Fundamentals of Wireless Communication*. Cambridge University Press, 2005.

[42] A. Marshall and I. Olkin, *Inequalities: Theory of Majorization and Its Applications*, ser. Mathematic in Science and Engineering. Academic Press, 1979, vol. 143.

[43] A. Wiesel, Y. C. Eldar, and S. Shamai (Shitz), "Zero-forcing precoding and generalized inverses," *IEEE Trans. on Sig. Proc.*, vol. 56, no. 9, pp. 4409–4418, Sept. 2008.

[44] H. Huh, H. C. Papadopoulos, and G. Caire, "MIMO broadcast channel optimization under general linear constraints," in *Proc. IEEE Int. Symp. on Inform. Theory (ISIT)*, Seoul, Korea, June 2009.

[45] R. Zhang, "Cooperative multi-cell block diagonalization with per-base-station power constraints," *IEEE J. Select. Areas Commun.*, vol. 28, no. 9, pp. 1435–1445, Dec. 2010.

[46] H. Shirani-Mehr and G. Caire, "Channel state feedback schemes for multiuser MIMO-OFDM downlink," *IEEE Trans. on Commun.*, vol. 57, no. 9, pp. 2713–2723, Sept. 2009.

[47] K. R. Kumar and G. Caire, "Channel state feedback over the MIMO-MAC," in *Proc. IEEE Int. Symp. on Inform. Theory (ISIT)*, Seoul, Korea, June 2009.

[48] M. Kobayashi, N. Jindal, and G. Caire, "Training and feedback optimization for multiuser MIMO downlink," *To appear in IEEE Trans. on Commun.*, 2011. [Online]. Available: http://arxiv.org/abs/0912.1987

[49] T. L. Marzetta and B. M. Hochwald, "Capacity of a mobile multiple-antenna communication link in Rayleigh flat fading," *IEEE Trans. on Inform. Theory*, vol. 45, no. 1, pp. 139–157, Jan. 1999.

[50] L. Zheng and D. N. C. Tse, "Communication on the Grassmann manifold: a geometric approach to the noncoherent multiple-antenna channel," *IEEE Trans. on Inform. Theory*, vol. 48, no. 2, pp. 359–383, Feb. 2002.

[51] B. Hassibi and B. M. Hochwald, "How much training is needed in multiple-antenna wireless links?" *IEEE Trans. on Inform. Theory*, vol. 49, no. 4, pp. 951–963, Apr. 2003.

[52] C. Wang, T. Gou, and S. A. Jafar, "Aiming perfectly in the dark–blind interference alignment through staggered antenna switching," *IEEE Trans. on Sig. Proc.*, vol. 59, no. 6, pp. 2734–2744, June 2011.

[53] A. M. Tulino and S. Verdu, *Random Matrix Theory and Wireless Communications*.   Foundations and Trends in Communications and Information Theory, 2004, vol. 1, no. 1.

[54] A. M. Tulino, A. Lozano, and S. Verdu, "Impact of antenna correlation on the capacity of multiantenna channels," *IEEE Trans. on Inform. Theory*, vol. 51, no. 7, pp. 2491–2509, July 2005.